\documentclass[9pt,a4paper,twocolumn]{IEEEtran}
\usepackage{graphicx}
\usepackage[comma,numbers,square,sort&compress]{natbib}
\usepackage{epstopdf}
\usepackage{amsmath}
\usepackage{enumerate}
\usepackage{subfigure}
\usepackage{ntheorem}
\usepackage{amssymb}
\usepackage{mathrsfs}        
\usepackage{indentfirst}

\newtheorem{lemma}{Lemma}

\begin{document}


\setlength{\footskip}{-1.0cm}
\addtolength{\footskip}{-1.20cm}

\title{ Parameter Recovery from Tangential Interpolations for Systems with an LFT Structure} 

\author{Tong Zhou and Yubing Li
\thanks{This work was supported in part by the NNSFC under Grant 62373212 and 62127809, and by the BNRist project under Grant BNR2024TD03003.}
\thanks{Tong Zhou and Yubing Li are with the Department of Automation, Tsinghua University, Beijing, 100084, P.~R.~China (email: {\tt\small tzhou@mail./liyb20@mails.tsinghua.edu.cn}). }}

\maketitle

\begin{abstract}       
This paper investigates how to recover parameters of a linear time invariant system from values and derivatives of its transfer function matrix, along several particular directions at a prescribed set of points in the complex plane, in which system matrices depend on these parameters through a linear fractional transformation. A necessary and sufficient condition is derived for a unique determination of these system parameters, which is expressed by a vector inequality. Under some particular situations, this condition reduces to a full column rank requirement on a constant matrix. Moreover, a method is given to recover system parameters from these values and derivatives, which is expressed by a vector linear equation with some rank constraints, for which various methods exist for finding its solutions. Robustness of the suggested recovery method is also clarified. A numerical example is given to illustrate characteristics of the suggested method, as well as effectiveness of derivative information introduction in parameter recovery, in which natural frequency and damping ratio are to be recovered for a transfer function.
\end{abstract}


\renewcommand{\labelenumi}{\rm\bf A\arabic{enumi})}

\addtolength{\abovedisplayskip}{-0.10cm}
\addtolength{\belowdisplayskip}{-0.10cm}

\addtolength{\abovedisplayshortskip}{-0.10cm}
\addtolength{\belowdisplayshortskip}{-0.10cm}

\setlength{\footskip}{-1.0cm}
\addtolength{\footskip}{-1.20cm}

\setlength{\parskip}{0.00cm}
\setlength{\itemsep}{0.0cm}
\setlength{\partopsep}{0.0cm}
\setlength{\topsep}{0.00cm}
\setlength{\leftmargin}{0.0cm}

\vspace{-0.25cm}
\section{Introduction}

\vspace{-0.25cm}
The value of a matrix valued function, as well as its derivatives, along a particular direction at a fixed point in the complex plane, is called its tangential interpolation. This value includes impulse response of a dynamic system, as well as its frequency response, as a special case. A long-standing and fundamental problem in systems and control theory, operator theory, functional analysis, etc. is to recover system matrices or transfer function matrix (TFM) of a linear time invariant (LTI) system from several tangential interpolations, that is, values of its TFM  and its TFM's derivatives at some prescribed places of the complex plane along several particular directions. Conclusions and methodologies developed for this problem play important roles in function reconstruction, system identification, model reduction and optimal control, etc. \cite{abg2020,adm1971,blgm1995,hk1966}.

Specifically, some well known results include the classical Carathedory-Fejer interpolation, Nevanlinna-Pick interpolation, which give a complete parametrization for all functions that are bounded over the closed unit disk of the complex plane and have some known values at several prescribed points inside this disk. These results lay foundations in developing theories for system realization, model reduction and robust control \cite{bgr1990,Glover1984,zdg1996, Zhou1998}, as well as estimation and validation of a model set that matches requirements for robust controller design \cite{pspi1998,Smith1995,zk1993}. Recently, it is observed in \cite{Astolfi2010} and \cite{Zhou2025} that each tangential interpolation of a TFM is related to the steady state response of the associated LTI system under the stimulation of a particular input signal. This observation enables extension of a moment matching based model reduction method for an LTI system to a nonlinear dynamic system, a time varying dynamic system, etc. \cite{Astolfi2010}, as well as tangential interpolation estimation and parameter identification from system output measurements with slow and non-periodic sampling \cite{Zhou2025}.

On the other hand, various results have been obtained for TFM/system matrix construction from tangential interpolations. For example, in \cite{hk1966}, a singular value decomposition (SVD) based procedure is suggested to construct a balanced realization of a LTI system from its impulse response, while a parametrization is given in \cite{blgm1995} for all systems that share the same covariance sequence. The so-called Loewner framework is adopted in \cite{abg2020} to reduce the complexity of a high dimensional descriptor form model through matching some of its left and right tangential interpolations. For a class of nonlinear descriptor systems satisfying some smoothness conditions and rank constraints, \cite{sma2024} gives an explicit expression for all systems that have a prescribed degree and meet several right tangential interpolation conditions. In all these works, coefficients of the TFM, or elements of the system matrices, are implicitly assumed to be algebraically independent of each other, meaning that system working principles have not been efficiently taken into account in building a system model. That is, the system is regarded as a black box.

In actual applications, however, a parameter usually has its own meanings in physics, chemistry, biology, economy, etc., or has some system significance in revealing response characteristics of a dynamic plant. For example, masses, springs, dampers in a mechanical system; substance concentrations, Michaelis-Menten constants in a gene regulation network; time constants, natural frequencies, damping ratios in a transfer function; subsystem interaction strengthes in a large scale system; etc. In system analysis and synthesis, the values of these parameters are widely regarded to be more helpful in various situations \cite{cw2020, gjmabchjkrs2019, Koopmans1949, pmssaxcas2010, Siljak1978}. When these factors must be taken into account in building a system model, neither its TFM coefficients nor its system matrix elements, can be regarded to be algebraically independent of each other \cite{mpr2014,zdg1996,zyl2018}. In addition, rather than to provide a set of possible values of a parameter in a system model, it is widely believed more helpful to give its accurate value, or an interval with a prescribed confidence, if this parameter is to be estimated from experimental data \cite{adm2020, tmbdms2015, Zhou2025}.

In describing relations between parameters of a system and coefficients of its TFM or elements of its system matrices, a linear fractional transformation (LFT) is widely adopted. For example, \cite{cw2020}, \cite{zdg1996} and \cite{zyl2018}. Note that any multi-variable rational function can be re-expressed by an LFT of its variables, and a cascade, parallel and feedback connection of any two LFTs with compatible dimensions can still be expressed by an LFT \cite{pd1993,Redheffer1960,zdg1996}. In addition, recent investigations show that system matrices of a networked dynamic system can also be expressed as an LFT of its subsystem connection parameters \cite{cw2020,zyl2018,Zhou2020-2}. It seems safe to declare that this LFT description is highly efficient and quite representative, and is capable of representing the aforementioned relations for a large class of LTI systems.

Motivated by these observations, this paper investigates possibilities and methods of uniquely recovering the values of system parameters from some right tangential interpolations of an LTI plant, in which system matrices depend on the parameters to be recovered through a known LFT. A necessary and sufficient condition has been derived for right tangential interpolations, that can uniquely determine the values of these parameters. This condition is in general in a vector inequality form, but can be reduced to a full column rank (FCR) constraint in some particular cases. In addition, on the basis of the parametrization of \cite{sma2024}, a method is suggested for this parameter recovery, in which a vector linear equation is to be solved under two rank constraints. Situations have also been clarified in which this recovery method is robust against estimation errors in a tangential interpolation.

The remaining of this paper is organized as follows. At first, in Section 2, problem descriptions and some preliminary results are given. Recovery possibility is attacked in Section 3, in which a necessary and sufficient condition is derived for uniquely determining the values of system parameters from right tangential interpolations. Section 4 investigates how to recover these parameter values with right tangential interpolations, while robustness against data inaccuracy is discussed in Section 5 for the suggested method. A numerical example is given in Section 6 to illustrate applicability and characteristics of these theoretical results, as well as effectiveness of derivative information introduction in improving parameter recovery accuracy. Some concluding remarks are given in Section 7 in which several further issues are discussed. Finally, two appendices are included. One gives proofs of some technical results, while the other provides system matrices for the simulation example.

The following notation and symbols are adopted in this paper. $\mathbb{R}^{m\times n}$ and $\mathbb{C}^{m\times n}$ stand respectively for the set of $m\times n$ dimensional real and complex matrices. When $m=1$ and/or  $n=1$, they are usually omitted, and the corresponding matrix  reduces to a row/column vector or scalar. $\bullet_{r}^\bot$/$\bullet_{l}^\bot$ represents a matrix whose columns/rows form a base of the right/left null space of a matrix, while ${\rm\bf row}\{X_i|^n_{i=1}\}$ the vector/matrix stacked horizontally by $X_i|^n_{i=1}$ with its $i$th column block vector/matrix being $X_i$, and ${\rm\bf vec}\{X\}$ the column vector stacked from left to right by the columns of the matrix $X$. The symbol $\otimes$ is used to denote the Kronecker product of two matrices, ${\rm\bf rank}(\bullet)$ the rank of a matrix, while  $\bullet^{\dag}$ the Moore-Penrose inverse of a matrix, $\bullet^{T}$ the transpose of a vector/matrix, and $||\bullet||$ a norm of a vector or a matrix. Specifically, $||\bullet||_{2}$ stands for the Euclidean norm of a vector or the Euclidean induced norm of a matrix, while $||\bullet||_{F}$ and $||\bullet||_{\star}$ respectively its Frobenius norm and nuclear norm. $\sigma_{i}\left( \bullet\right)$ stands for the $i$th singular value of a matrix arranging as $\sigma_{1}\left( \bullet\right) \geq \sigma_{2}\left( \bullet\right) \geq \cdots \geq 0$, while $I_{m}$ and $0_{m\times n}$ respectively for the $m$ dimensional identity matrix and the $m\times n$ dimensional zero matrix. The subscripts are often omitted when this piece of information is not very essential. $o\!\left( x_{1},x_{2},\cdots,x_{m}\right)$ with $x_{i}\geq 0$ for each $i=1,2,\cdots,m$, represents a higher order infinitesimal of $\max_{1\leq i \leq m} x_{i}$.

\section{Problem Formulation and Preliminaries}\label{section:pfp}

Consider an LTI continuous-time system $\mathbf{\Sigma}_{p}$ that has $m_{\theta}$ parameters $\theta_{i} | _{i=1}^{m_{\theta}}$, and is widely adopted in dynamic system analysis and synthesis, for example, \cite{pd1993}, \cite{Smith1995}, \cite{zdg1996} and \cite{zyl2018}, etc. More specifically, define a vector $\theta$ as $\theta = col \left\lbrace \theta_{i} | _{i=1}^{m_{\theta}} \right\rbrace $, and assume that the input-output relations of this system $\mathbf{\Sigma}_{p}$ can be described as
\begin{eqnarray}
	\dot{x}(t)  &=& A(\theta)x(t)+B(\theta)u(t)\label{plant-1}\\
	y(t) &=& C(\theta)x(t)+D(\theta)u(t)   \label{plant-2}
\end{eqnarray}
\hspace*{-0.00cm}Here, $x(t)\in\mathbb{R}^{m_{x}}$, $u(t)\in\mathbb{R}^{m_{u}}$ and $y(t)\in\mathbb{R}^{m_{y}}$ are respectively the system state, input and output vectors. Moreover, assume that the system matrices $A(\theta)$, $B(\theta)$, $C(\theta)$ and $D(\theta)$ depend on its parameters through the following LFT form,
\begin{eqnarray}\label{plant-3}
\hspace*{-0.5cm} \left[\!\!\begin{array}{cc}
				A(\theta) & B(\theta)\\
				C(\theta) & D(\theta)
		\end{array}\!\!\right] \!\!
&=& \!\!
		\left[\!\!\begin{array}{cc}A_{xx} & B_{xu}\\
				C_{yx} & D_{yu}\\
		\end{array}\!\!\right] +
        \left[\!\begin{array}{c}
				B_{xv}\\
				D_{yv}
        \end{array}\!\right]\!
        \left[ I_{m_{v}} \!\!-\!\! P(\theta)D_{zv}\right]^{-\!1}\!\times \nonumber\\
		& &\hspace*{2cm}  P(\theta)
        \left[\begin{array}{cc}
				C_{zx} & D_{zu}
        \end{array}\right]
\end{eqnarray}
\hspace*{-0.00cm}in which
\begin{equation}\label{plant-4}
	P(\theta)= P_{0} + \sum\limits_{i=1}^{m_{\theta}}\theta_{i}P_{i}
\end{equation}
\hspace*{-0.00cm}It is also assumed that a set $\mathbf{\Theta}$ is given that consists of all possible values of the parameter vector $\theta$, which is determined by some a priori plant information. In addition, assume that $A_{xx}\in \mathbb{R}^{m_{x}\times m_{x}},\; B_{xu}\in \mathbb{R}^{m_{x}\times m_{u}},\; B_{xv}\in \mathbb{R}^{m_{x}\times m_{v}},\; C_{yx}\in \mathbb{R}^{m_{y}\times m_{x}},\; C_{zx}\in \mathbb{R}^{m_{z}\times m_{x}},\; D_{zu}\in \mathbb{R}^{m_{z}\times m_{u}},\; D_{zv}\in \mathbb{R}^{m_{z}\times m_{v}},\; D_{yu}\in \mathbb{R}^{m_{y}\times m_{u}}$ and $D_{yv}\in \mathbb{R}^{m_{y}\times m_{v}}$, as well as  $P_{i}\in \mathbb{R}^{m_{v}\times m_{z}}$ with $i=0, 1, \cdots, m_{\theta}$, are some prescribed matrices, reflecting available a priori knowledge about the system from its working principles, or some particular forms describing its input-output relations that are convenient in system analysis and synthesis, etc.

\newtheorem{Assumption}{Assumption}
To make the problem to be discussed meaningful, it is assumed throughout this paper that System $\mathbf{\Sigma}_{p}$ is well-posed for each $\theta \in \mathbf{\Theta}$. That is, the following assumption is adopted.

\begin{Assumption} \label{assum:1}
For each $\theta \in \mathbf{\Theta}$, the matrix $I_{m_{v}} \!\!-\!\! P(\theta)D_{zv}$ is invertible. In addition, the set $\mathbf{\Theta}$ is compact, meaning it is bounded and closed.
\end{Assumption}

Let $H(s,\theta)$ denote the TFM of System $\mathbf{\Sigma}_{p}$. A right tangential interpolation of this system is defined as follows, with respect to a particular location in the complex plane, and a prescribed direction \cite{abg2020, bgr1990, Zhou2025}.

\newtheorem{Definition}{Definition}

\begin{Definition}\label{def:1}
Let $\eta \in \mathbb{C}^{m_{u}}$ be an arbitrary complex vector, while $\lambda \in \mathbb{C}$ be an arbitrary complex number. For each non-negative integer $k$, denote the $k$th derivative of the TFM $H(s,\theta)$ at $\lambda$ by ${d^{k}H(\lambda,\theta)}/{ds^{k}}$. Then
\begin{displaymath}
\zeta = \frac{d^{k}H(\lambda,\theta)}{ds^{k}}\eta
\end{displaymath}
is called the right tangential interpolation of System $\mathbf{\Sigma}_{p}$ at $\lambda$ along the direction $\eta$.
\end{Definition}

Let $\Xi \in \mathbb{R}^{m_{\xi}\times m_{\xi}}$ and $\Pi \in \mathbb{R}^{m_{u}\times m_{\xi}}$ be two prescribed real matrices. Assume that the matrices $A(\theta)$ and $\Xi$ do not share any eigenvalues. Then it can be declared from matrix theories \cite{gv1989, hj1991} that, the following matrix equation has a unique solution,
\begin{equation}\label{tan-int-1}
	A(\theta)X + B(\theta)\Pi - X\Xi = 0 	
\end{equation}
Define a matrix $\Gamma$ as
\begin{equation}\label{tan-int-2}
	\Gamma = C(\theta)X + D(\theta)\Pi  	
\end{equation}
It has been shown in \cite{Zhou2025} that there is a bijective map between this matrix and a linear combination of some right tangential interpolations of System $\mathbf{\Sigma}_{p}$ with constant coefficients, and these coefficients depend on neither the system matrices of System $\mathbf{\Sigma}_{p}$, nor matrices $\Xi$ and $\Pi$. The locations of the associated right tangential interpolations are completely determined by the eigenvalues of the matrix $\Xi$, and the maximum degree of the derivatives of the TFM $H(s,\theta)$ at an interpolation point by the geometric multiplicity of the associated eigenvalue, while the interpolation directions by the matrix $\Pi$ and a transformation matrix of the matrix $\Xi$ that gives its Jordan canonical form.

In addition, \cite{Zhou2025} has also clarified that the matrix $\Gamma$ of Equation (\ref{tan-int-2}) can be efficiently estimated from measured outputs of System $\mathbf{\Sigma}_{p}$ under a particular stimulation, in which sampling is allowed to be non-uniform and slower than the Nyquist frequency. More precisely, assume that the input signal $u(t)$ of this system is generated by the following autonomous LTI system $\mathbf{\Sigma}_{s}$, with its state vector $\xi(t)$ belonging to $\mathbb{R}^{m_{\xi}}$,
\begin{displaymath}
	\dot{\xi}(t)  = \Xi \xi(t), \hspace{0.5cm}
	u(t) = \Pi \xi(t)  \label{signal-1}
\end{displaymath}
Then Theorem 1 of \cite{Zhou2025} declares that the steady state response of System $\mathbf{\Sigma}_{p}$ depends linearly on the aforementioned matrix $\Gamma$. This observation enables its estimation from experimental data with some well established methods, such as least squares estimation, maximum likelihood estimation, etc., provided that System $\mathbf{\Sigma}_{p}$ is stable.

Similarly, a left tangential interpolation can also be defined for this system. However, rather than the output of System $\mathbf{\Sigma}_{p}$, this interpolation is more closely and directly related to a filtered system output. It is therefore not adopted in this investigation. Nevertheless, the results of this study can be directly extended to left tangential interpolations, simply taking transpose of their counterparts. Based on these observations, a right tangential interpolation is sometimes also simply called a tangential interpolation in this paper for brevity.

From the above arguments and observations, the next definition comes naturally.

\begin{Definition}\label{def:2}
For each matrix pair $\Xi \in \mathbb{R}^{m_{\xi}\times m_{\xi}}$ and $\Pi \in \mathbb{R}^{m_{u}\times m_{\xi}}$, the matrix $\Gamma$ defined by Equations (\ref{tan-int-1}) and (\ref{tan-int-2}) is called the right tangential interpolation matrix (RTIM) of System $\mathbf{\Sigma}_{p}$ at the matrix $\Xi$ along the directions of the matrix $\Pi$, or simply at the matrix pair $(\Xi,\;\Pi)$. In addition, the value of the parameter vector $\theta$, or simply the parameter vector $\theta$, of System $\mathbf{\Sigma}_{p}$ is called (globally) recoverable from its RTIM $\Gamma$ at $(\Xi,\;\Pi)$, if this value can be uniquely determined from this given RTIM.
\end{Definition}

In \cite{Astolfi2010} and \cite{sma2024}, the matrix $\Gamma$ is also called moments of System $\mathbf{\Sigma}_{p}$ at the matrix pair $(\Xi,\;\Pi)$. This is quite perfect for a single-input single-output system, noting that an interpolation direction is not very essential in the case. For a multi-input multi-output system, however, different interpolation directions generally carry different information about the system. To emphasize this difference, the above definition is adopted in this paper.

With these preparations, the problem investigated in this paper can be formulated as follows.

\hspace*{-0.40cm}{\bf Problem.} {\it Given a RTIM $\Gamma$ of System $\mathbf{\Sigma}_{p}$, as well as the associated matrices $\Xi$ and $\Pi$, determine whether the value of its parameter vector $\theta$  is recoverable from this system information. If the answer is positive, develop a method to recover this value.}

When a RTIM $\Gamma$ is available, \cite{sma2024} gives a complete parametrization for all systems that match this matrix and have a degree not smaller than $m_{\xi}$. This parametrization plays important roles for recovering the value of the parameter vector $\theta$ of System $\mathbf{\Sigma}_{p}$ from a RTIM, and is therefore restated in the next lemma.

\newtheorem{Lemma}{Lemma}

\begin{lemma}
Assume that $\Xi \in \mathbb{R}^{m_{\xi}\times m_{\xi}}$ and $\Pi \in \mathbb{R}^{m_{u}\times m_{\xi}}$ be two prescribed matrices. Moreover, assume that $\Gamma \in \mathbb{R}^{m_{y}\times m_{\xi}}$ is a prescribed RTIM. Then for each positive integer $m_{x}$ satisfying $m_{x} \geq  m_{\xi}$, all the LTI systems that have a $m_{x}$ dimensional state vector and match the RTIM $\Gamma$, can be parameterized by the following state-space realization, modulo a nonsingular state transformation,
\begin{eqnarray*}
	\dot{x}(t)  &=& \left[\begin{array}{cc} \Xi - G \Pi  & Z \\ -S \Pi  &  F \end{array}\right] x(t) +
\left[\begin{array}{c}  G  \\ S  \end{array}\right] u(t)  \\
	y(t) &=& \left[\begin{array}{cc} \Gamma - K \Pi  & H \end{array}\right] x(t)+ K u(t)
\end{eqnarray*}
in which $G \in \mathbb{R}^{m_{\xi}\times m_{u}}$, $Z \in \mathbb{R}^{m_{\xi}\times (m_{x}-m_{\xi})}$, $S \in \mathbb{R}^{(m_{x}-m_{\xi})\times m_{u}}$, $F \in \mathbb{R}^{(m_{x}-m_{\xi})\times (m_{x}-m_{\xi})}$, $ K \in \mathbb{R}^{m_{y}\times m_{u}}$ and $H \in \mathbb{R}^{m_{y}\times (m_{x}-m_{\xi})}$ are arbitrary real matrices.
\label{lemma:1}
\end{lemma}

To derive a computationally feasible condition for verifying recoverability of the parameter vector $\theta$ for System $\mathbf{\Sigma}_{p}$ from a RTIM $\Gamma$, as well as to develop a computationally feasible procedure to recover its value, the following results are introduced \cite{gv1989, hj1991, zly2024, zyl2018}.

\begin{lemma}
Assume that a matrix $A$ is partitioned as $A = \left[ A_{1}^{T} \;  A_{2}^{T}\right]^{T}$. Moreover, assume that the matrix $A_{1}$ is not of FCR. Then the matrix $A$ is of FCR, if and only if the matrix $A_{2}A_{1,r}^{\perp}$ is.
\label{lemma:2}
\end{lemma}

When the matrix $A_{1}$ is of FCR, $A_{1,r}^{\perp}=0$. In this case, for an arbitrary matrix $A_{2}$ with the same number of columns as the matrix $A_{1}$, the matrix $A = \left[ A_{1}^{T} \;  A_{2}^{T}\right]^{T}$ is obviously always of FCR, meaning that the above lemma is not valid in this case.

\begin{lemma}
Let $A$, $B$ and $C$ be some prescribed real matrices with compatible dimensions. There exists a matrix $X$ satisfying $AXB = C$, if and only if the following two equalities are satisfied simultaneously
\begin{displaymath}
A_{l}^{\perp}C = 0, \hspace{0.5cm} CB_{r}^{\perp}=0
\end{displaymath}
In addition, under the satisfaction of these two conditions, all the matrices satisfying $AXB = C$ can be explicitly parameterized by the following formula,
\begin{displaymath}
X = A^{\dag} C B^{\dag} + Z -A^{\dag}A Z BB^{\dag}
\end{displaymath}
in which $Z$ is an arbitrary real matrix that has an appropriate dimension.
\label{lemma:3}
\end{lemma}

The following results give a necessary condition for uniquely determining the value of the parameter vector $\theta$ of System $\mathbf{\Sigma}_{p}$ from its RTIM $\Gamma$, noting that identifiability is always a prerequisite for parameter estimations and recovery \cite{adm2020,zly2024}.

\begin{lemma}\label{lemma:4}
Define a matrix $\Psi$ using the matrices $P_{i}|_{i=1}^{m_{\theta}}$ of Equation (\ref{plant-4}) as follows,
\begin{equation}
\Psi = \left[ {\rm\bf vec}(P_{1})\;\; {\rm\bf vec}(P_{2})\;\; \cdots \;\; {\rm\bf vec}(P_{m_{\theta}}) \right]
\label{eqn:par-vec-1}
\end{equation}
Then the parameter vector $\theta$ of System $\mathbf{\Sigma}_{p}$ is globally identifiable, only when the matrix $\Psi$ is of FCR.
\end{lemma}

\section{Recoverability of the Parameter Vector}

In this section, recoverability is discussed for the value of the parameter vector $\theta$ of System $\mathbf{\Sigma}_{p}$ from its RTIM $\Gamma$, that is, conditions for uniquely determining this value from some information about its right tangential interpolations. To clarify local and global characteristics of the derived conditions, a solution $X$ to Equation (\ref{tan-int-1}) is rewritten as $X(\theta)$, while the associated RTIM $\Gamma$ of Equation (\ref{tan-int-2}) as $\Gamma(\theta)$, and $\mathbf{\Sigma}_{p}$ as  $\mathbf{\Sigma}_{p}(\theta)$, explicitly expressing their dependence on the value to be recovered.

\newtheorem{Theorem}{Theorem}

\begin{Theorem}\label{theo:1}
Assume that System $\mathbf{\Sigma}_{p}(\theta)$ is well-posed, and the matrix $\Psi$ defined in Lemma \ref{lemma:4} is of FCR. Moreover, assume that Equation (\ref{tan-int-1}) has a solution $X(\theta)$. Then the value of the parameter vector $\theta$ of System $\mathbf{\Sigma}_{p}(\theta)$ can be uniquely determined by its RTIM $\Gamma(\theta)$, if and only if for each nonzero real vector $\phi$ having a compatible dimension, the following inequality is valid.
{\small \begin{eqnarray}
& & \hspace*{-1.0cm}\!\!
  \left\{\left(\left\{\left[C_{zx} X(\theta) + D_{zu} \Pi\right]^{T}\left[ I-D_{zv} P(\theta)\right]^{-T}\right\}\otimes I\right) \Psi  -
  \right. \nonumber \\
& & \hspace*{-1.0cm}\!\!
  \left.{\rm\bf row} \!\!\left(\!\! \left.
  \left[\! I \!\!\otimes\!\!\! \left( \!\!P_{i}D_{zv} \!\!\left[\!\!\!\!\begin{array}{c}
				B_{xv}\\
				D_{yv}
        \end{array}\!\!\!\!\right]_{r}^{\perp} \!\!\right) \!\!\right] \!\!\phi\right|_{i=1}^{m_{\theta}} \!\right) \!\!\!\right\}_{\!l}^{\!\perp} \!\!\!\!\!
  \left( \!\! I \!\!\otimes\!\! \left\{\!\! \left[ I \!-\! P(\theta)D_{zv} \right] \!\!
  \left[\!\!\!\!\begin{array}{c}
				B_{xv}\\
				D_{yv}
        \end{array}\!\!\!\!\right]_{r}^{\perp} \!\!\!\right\} \!\!\right) \!\!\phi \!\neq \! 0
\label{eqn:theo:1}
\end{eqnarray}}
\end{Theorem}

A proof of this theorem is given in Appendix I.

When the matrices $A(\theta)$ and $\Xi$ do not share an eigenvalue, \cite{Zhou2025} shows that the matrix $X(\theta)$ can be expressed as an LFT of the parameter vector $\theta$. That is, each element of this matrix is a proper rational function of this vector. From this conclusion, as well as the definition of the matrix $P(\theta)$, it is obvious that each element of the matrix $\left[C_{zx} X(\theta) + D_{zu} \Pi\right]^{T}\left[ I-D_{zv} P(\theta)\right]^{-T}$ is also a proper rational function of the variables $\theta_{i}|_{i=1}^{m_{\theta}}$.
Note that each element in a base of the left/right null space of a matrix with rational function elements is still a rational function \cite{gv1989,hj1991,Khazanov2006}. It can be declared that each element of the vector in the left side of Equation (\ref{eqn:theo:1}) is a rational function of the vectors $\theta$ and $\phi$, meaning that the conditions of Theorem \ref{theo:1} can in principle be verified through symbolic computations \cite{besd2021,Khazanov2006}. This verification, however, is currently still quite time and memory consuming in general, especially when the dimension of System $\mathbf{\Sigma}_{p}(\theta)$, or the vector $\theta$/$\phi$, is large.

Note that a real vector is not equal to zero, if and only if the product of its transpose and itself is greater than zero. As each element of the vector in the left side of Equation (\ref{eqn:theo:1}) is a rational function, it is obvious that the product of its transpose and itself is also a rational function. Hence, this product is analytic at almost every feasible value of the vectors $\theta$ and $\phi$, meaning that its value with a nonzero $\phi$ has only three possibilities, that is, constantly equal to zero, equal to zero in a set with a zero measure, always greater than zero. From these observations, a pragmatic method for verifying the conditions of Theorem \ref{theo:1} appears to be a repetition of the following three-step procedure with an appropriate number of times, say $N_{\theta}$.

\begin{itemize}
\setlength{\parskip}{0.0cm}
\setlength{\itemsep}{0.0cm}
\setlength{\partopsep}{0.0cm}
\setlength{\topsep}{0.0cm}
\setlength{\leftmargin}{0.0cm}
\item randomly generate a value of the vector $\theta$ in the set $\mathbf{\Theta}$, according to a continuous uniform distribution.
\begin{itemize}
\item randomly generate a value of the vector $\phi$ in a bounded subset of $\mathbb{R}^{m_{\phi}}$, according to a continuous uniform distribution. Here, $m_{\phi}$ stands for the dimension of the vector $\phi$.
\item compute the product of the transpose of the vector in the left side of Equation (\ref{eqn:theo:1}) and itself with the generated values for the vector pair $(\theta,\;\phi)$.
\item verify whether the computed value is greater than $\mu_{t}\phi^{T}\phi$ for a prescribed threshold parameter $\mu_{t}$.
\end{itemize}
\item repeat the above three steps to a prescribed number of times, say $N_{\phi}$.
\end{itemize}

If the number of the generated samples, that is, $N_{\theta}$ or $N_{\phi}$, is sufficiently large for both the vector $\theta$ and the vector $\phi$, and all the generated samples passed the above test, then it is with a high probability that each value of the parameter vector $\theta$, that is, each element of the set $\mathbf{\Theta}$, satisfies the conditions of Theorem \ref{theo:1}, and is therefore globally recoverable from a RTIM $\Gamma(\theta)$ of System $\mathbf{\Sigma}_{p}(\theta)$, meaning that this system is globally recoverable. In other words, parameter recoverability is a global property of System $\mathbf{\Sigma}_{p}(\theta)$ in a high probability. On the other hand, if some of the generated samples for the vector $\theta$ do not pass this test, then this parameter vector is globally recoverable only for values belonging to a strict subset of the set $\mathbf{\Theta}$, and parameter recoverability is a local property of System $\mathbf{\Sigma}_{p}(\theta)$. Otherwise, this parameter vector is not globally recoverable at each element of the set $\mathbf{\Theta}$.

The above procedure is relatively standard in verifying a system property through random samples, and some rigorous conclusions have been established for the number of samples to achieve a prescribed confidence level of the verification result \cite{cdt2011}. These results can be borrowed for establishing relations between the sample numbers $N_{\theta}$ and  $N_{\phi}$ and the confidence level of the conclusions about parameter recoverability of System $\mathbf{\Sigma}_{p}(\theta)$. In addition, this procedure can also be adopted in the verification of Assumption \ref{assum:1}, noting that a square matrix is regular, if and only if it is of FCR. The details are omitted due to space considerations, as well as that this is not a major issue to be discussed in this paper.

On the other hand, Equation (\ref{eqn-app-19}) in the proof of Theorem \ref{theo:1} reduces to a linear matrix equation under some particular cases. This reduction makes the conditions of Theorem \ref{theo:1} become a FCR requirement on a constant matrix depending only on the value of the parameter vector $\theta$. The following corollary gives two situations, while its proof is deferred to Appendix I.

\newtheorem{Corollary}{Corollary}

\begin{Corollary} \label{coro:1}
Assume that all the three hypotheses of Theorem \ref{theo:1} are satisfied.
\begin{itemize}
\item Assume further that $\left[ B^{T}_{xv} \;\; D^{T}_{yv} \right]_{l}^{\perp} D^{T}_{zv} = 0$. Then the parameter vector $\theta$ is recoverable from the RTIM $\Gamma(\theta)$, if and only if the following matrix is of full row rank (FRR),
\begin{equation}
 \hspace*{-1.2cm}
\left[\!\!\! \begin{array}{c}
\Psi^{T}
\left(\left\{\left[ I-D_{zv} P(\theta)\right]^{-1}\left[C_{zx} X(\theta) + D_{zu} \Pi\right]\right\}\otimes I\right)  \\
 I \otimes
\left[ B^{T}_{xv} \;\; D^{T}_{yv}
\right]_{l}^{\perp}   \end{array} \!\!\!\right]
\label{eqn:coro:1-1}
\end{equation}

\item Assume further that the matrix $C_{zx} X(\theta) + D_{zu} \Pi$ is of FRR. Then the parameter vector $\theta$ is recoverable from the RTIM $\Gamma(\theta)$, if and only if the following matrix is of FCR,
\begin{equation}
 \hspace*{-1.2cm}
\left[\!\! \begin{array}{c} I \otimes \left[ B^{T}_{xv}\;\; D^{T}_{yv} \right]^{T}\\
        \Psi_{l}^{\perp}\left(I \otimes \left[ I - P(\theta)D_{zv} \right]\right)  \end{array} \!\!\right]
\label{eqn:coro:1-2}
\end{equation}
\end{itemize}
\end{Corollary}

In addition, some necessary conditions can also be derived from Theorem \ref{theo:1} for parameter vector recoverability of System $\mathbf{\Sigma}_{p}$, that can be verified without significant difficulties.

\begin{Corollary} \label{coro:2}
Under the same assumptions of Theorem \ref{theo:1}, the value of the parameter vector $\theta$ of System $\mathbf{\Sigma}_{p}$ can be uniquely determined from its RTIM $\Gamma(\theta)$, only if the following two conditions are satisfied.
\begin{itemize}

\item For each nonzero real vector $\phi$ with an appropriate dimension, the following matrix is not of FRR,
\begin{eqnarray}
& & \hspace*{-1.2cm}\!\!
  \left(\left\{\left[ I-D_{zv} P(\theta)\right]^{-1}\left[C_{zx} X(\theta) + D_{zu} \Pi\right]\right\}^{T}\otimes I\right) \Psi  -
   \nonumber \\
& & \hspace*{2.2cm}\!\!
  {\rm\bf row} \!\left(\! \left.
  \left[I \!\otimes\! \left( \!P_{i}D_{zv} \!\left[\!\!\!\!\begin{array}{c}
				B_{xv}\\
				D_{yv}
        \end{array}\!\!\!\!\right]_{r}^{\perp} \!\right) \!\right] \phi\right|_{i=1}^{m_{\theta}} \!\right)
\label{eqn:coro:2-1}
\end{eqnarray}

\item The following matrix is of FCR,
\begin{equation}
\left[\!\! \begin{array}{c} \left[C_{zx} X(\theta) + D_{zu} \Pi\right]^{T} \otimes I \\
 \Psi_{l}^{\perp} \left(\left[ I-D_{zv} P(\theta)\right]^{T} \otimes I\right) \end{array} \!\! \right]
\label{eqn:coro:2-2}
\end{equation}
\end{itemize}
\end{Corollary}

A proof of this corollary is given in Appendix I.

Note that the matrix of Equation (\ref{eqn:coro:2-4}) is of FCR, only when the matrix $\Psi$ is of FCR. The proof of Corollary \ref{coro:2} also confirms necessity of the condition of Lemma \ref{lemma:4} from another aspect.

It is worthwhile to mention that each matrix in Corollaries \ref{coro:1} and \ref{coro:2} depends on the parameter vector $\theta$, meaning that verification of the associated condition requires knowledge about the unknown value of this parameter vector, which is not an appreciative characteristic in actual applications.

Nevertheless, when the matrices $A(\theta)$ and $\Xi$ do not have a common eigenvalue, each element of the matrix $X(\theta)$ is a rational function of the parameter vector $\theta$. It can therefore be declared directly from the definition of the matrix $P(\theta)$ that, every element of the four matrices in Equations (\ref{eqn:coro:1-1})--(\ref{eqn:coro:2-2}) is also a rational function of the vectors $\theta$ and $\phi$. For a matrix with each of its elements being a rational function of some variables, a FCR/FRR property is a generic property, meaning that this property does not depend on an actual value of the variables almost surely. More specifically, assume that each element of a matrix $\Lambda(\theta, \phi)$ is a rational function of the vectors $\theta$ and $\phi$. Then ${\rm\bf det}(\Lambda^{T}(\theta, \phi)\Lambda(\theta, \phi))$ is also a rational function of these two vectors, in which ${\rm\bf det}(\star)$ stands for the determinant of a square matrix. Note that a matrix  is of FCR, if and only if the determinant of the product of its transpose and itself is greater than zero \cite{gv1989,hj1991}. Moreover, a rational function with a finite degree is equal to zero, only when it is constantly equal to zero, or it is equal to zero only in a zero-measure set of its variables. Hence, when a FCR property is to be investigated for the matrix $\Lambda(\theta, \phi)$, only two cases exist. One case is that it is not FCR for each value of these two vectors, while the other is that it is almost surely of FCR. Similar arguments are valid for its FRR property. These observations imply that conditions of Corollaries \ref{coro:1} and \ref{coro:2} can be verified in a pragmatic way through simply sampling over the set $\mathbf{\Theta}$ and a bounded subset of $\mathbb{R}^{m_{\phi}}$ randomly, and checking the associated conditions with these samples.

\section{Parameter Recovery from RTIM}

To recover the value of the parameter vector $\theta$ of System $\mathbf{\Sigma}_{p}$ from a given RTIM $\Gamma$, a necessary and sufficient condition is at first established for the consistency of this value with the given information about right tangential interpolations of this system, which is given by the following theorem, and is derived from the parametrization of \cite{sma2024}, that is restated as Lemma \ref{lemma:1} in this paper. Its proof can be found in Appendix I.

\begin{Theorem}\label{theo:2}
Assume that $m_{x}\geq m_{\xi}$ and a RTIM $\Gamma$ is available for System $\mathbf{\Sigma}_{p}$ at the matrix $\Xi$ along the directions of the matrix $\Pi$. Then a value of the parameter vector $\theta$ is consistent with these right tangential interpolations, if and only if the following equation has a solution $T_{1}$ that is of FCR,
\begin{equation}
\left[\!\!\begin{array}{cc}
				A(\theta) & B(\theta)\\
				C(\theta) & D(\theta)
		\end{array}\!\!\right] \left[\!\!\begin{array}{c}
				T_{1} \\
				\Pi
		\end{array}\!\!\right]
= \left[\!\!\begin{array}{c}
				T_{1}\Xi \\
				\Gamma
		\end{array}\!\!\right]
\label{eqn:pr-1}
\end{equation}
\end{Theorem}

From Definition \ref{def:2}, it is obvious that for any value of the parameter vector $\theta$, if there is a matrix $T_{1}$ such that Equation (\ref{eqn:pr-1}) is satisfied, then this parameter vector value is certainly consistent with the right tangential interpolation information contained in the RTIM $\Gamma$, even if this matrix $T_{1}$ is not of FCR. On the other hand, note that when
the eigenvalues of the matrix $A(\theta)$ do not have any overlaps with those of the matrix $\Xi$, Equation (\ref{tan-int-1}) has a unique solution. It can therefore be declared that in this case, the solution $X$ is certainly of FCR.

The above theorem makes it clear that any value of the parameter vector $\theta$ in the set $\mathbf{\Theta}$ can not be invalidated by the given RTIM $\Gamma$ of System $\mathbf{\Sigma}_{p}$, provided that it satisfies Equation (\ref{eqn:pr-1}). In other words, to completely recover the value of this parameter vector from some information about right tangential interpolations of the system $\mathbf{\Sigma}_{p}$, it is essential that Equation (\ref{eqn:pr-1}) has a unique solution in the set $\mathbf{\Theta}$, meaning that the conditions of Theorem \ref{theo:1} must be satisfied.

In case that the dimension $m_{\xi}$ of the matrix $\Xi$ is greater than the dimension $m_{x}$ of System $\mathbf{\Sigma}_{p}$, the above theorem is still applicable, through simply transforming the matrix $\Xi$ into a block diagonal one.

More precisely, assume that there exists a nonsingular matrix $T_{\xi} \in \mathbb{R}^{m_{\xi}\times m_{\xi}}$ such that $\Xi = T_{\xi} {\rm\bf diag}\{\Xi_{i}|^{n_{\xi}}_{i=1}\} T^{-1}_{\xi}$, in which $\Xi_{i} \in \mathbb{R}^{m_{\xi,i}\times m_{\xi,i}}$ and $m_{\xi,i} \leq m_{x}$ for each $i=1,2,\cdots,n_{\xi}$, while
${\rm\bf diag}\{\Xi_{i}|_{i=1}^{n_{\xi}}\}$ stands for a block diagonal matrix with its $i$th diagonal block being $\Xi_i$. Partition the matrix $T_{\xi}$ as $T_{\xi}={\rm\bf row}\{T_{\xi,i}|_{i=1}^{n_{\xi}}\}$ consistently in dimensions with the matrices $\Xi_{i}|_{i=1}^{n_{\xi}}$. Let $X \in \mathbb{R}^{m_{x}\times m_{\xi}}$ and $\Gamma \in \mathbb{R}^{m_{y}\times m_{\xi}}$ be two arbitrary matrices. Define matrices $\overline{X}_{i}|_{i=1}^{n_{\xi}}$,  $\overline{\Pi}_{i}|_{i=1}^{n_{\xi}}$ and $\overline{\Gamma}_{i}|_{i=1}^{n_{\xi}}$ respectively as
\begin{displaymath}
\overline{X}_{i} = X T_{\xi,i}, \hspace{0.5cm}
\overline{\Pi}_{i} = \Pi T_{\xi,i}, \hspace{0.5cm}
\overline{\Gamma}_{i} = \Gamma T_{\xi,i}
\end{displaymath}
Then straightforward algebraic manipulations show that, the matrix pair $(X,\Gamma)$ satisfies Equations (\ref{tan-int-1}) and (\ref{tan-int-2}), if and only if for each $i=1,2,\cdots,n_{\xi}$, the following two equalities are satisfied simultaneously,
\begin{equation}
	A(\theta)\overline{X}_{i} + B(\theta)\overline{\Pi}_{i} - \overline{X}_{i}\Xi_{i} = 0, \hspace{0.50cm}	
	\overline{\Gamma}_{i} = C(\theta)\overline{X}_{i} + D(\theta)\overline{\Pi}_{i}  	
\end{equation}
meaning that if $\Gamma$ is a RTIM of System $\mathbf{\Sigma}_{p}$, then for each $1 \leq i \leq n_{\xi}$, $\overline{\Gamma}_{i}$ can be regarded to be its RTIM at the matrix pair $(\Xi_{i},\overline{\Pi}_{i})$. It can therefore be declared from Theorem \ref{theo:2} that, a value of the parameter vector $\theta$ is consistent with a RTIM $\Gamma$ of System $\mathbf{\Sigma}_{p}$ at the matrix pair $(\Xi,\Pi)$, if and only if for each $i=1,2,\cdots,n_{\xi}$, there exists a FCR $T_{1,i}$ satisfying
\begin{equation}
\left[\!\!\begin{array}{cc}
				A(\theta) & B(\theta)\\
				C(\theta) & D(\theta)
		\end{array}\!\!\right] \left[\!\!\begin{array}{c}
				T_{1,i} \\
				\overline{\Pi}_{i}
		\end{array}\!\!\right]
= \left[\!\!\begin{array}{c}
				T_{1,i}\Xi_{i} \\
				\overline{\Gamma}_{i}
		\end{array}\!\!\right]
\end{equation}

For brevity and space considerations, parameter recovery is not discussed in this paper in case that $m_{x} \leq m_{\xi}$. Nevertheless, the above arguments reveal that all the following results can be directly extended to this case, noting that the above transformation for the matrix $\Xi$ is always feasible \cite{gv1989,hj1991}.

To solve Equation (\ref{eqn:pr-1}), define matrices $\Upsilon_{t}$ and $\Upsilon_{s}$ respectively as follows,
{\small
\begin{equation}
\hspace*{-0.00cm}
\Upsilon_{t} \!=\!  I_{m_{\xi}} \!\otimes\!\!\ \left[\!\!\!\! \begin{array}{c}
				I_{m_{x}}\\
				0_{m_{y}\times m_{x}}
        \end{array}\!\!\!\! \right] \!, \hspace{0.10cm}
\Upsilon_{s} \!=\!
 \left(\!\Xi^{T} \!\!\otimes\!\! \left[\!\!\! \begin{array}{c}
				B_{xv}\\
				D_{yv}
        \end{array}\!\!\! \right]_{l}^{\perp} \!\!\right) \!\Upsilon_{t} \!-\! I \otimes \left(\!
\left[\!\!\! \begin{array}{c}
				B_{xv}\\
				D_{yv}
        \end{array}\!\!\! \right]_{l}^{\perp} \!\!
\left[\!\!\! \begin{array}{c} A_{xx} \\
				C_{yx} \\
		\end{array}\!\!\! \right]\!\right)
\label{eqn:pr-29}
\end{equation}}
Moreover, partition both the matrix $\Upsilon_{s,r}^{\perp}$ and the vector $\Upsilon_{s}^{\dag} {\rm\bf vec}\left(\left[\!\!\begin{array}{c}
				B_{xv}\\
				D_{yv}
        \end{array}\!\! \right]_{l}^{\perp} \!
\left[\!\!\begin{array}{c }  B_{xu} \Pi  \\
				D_{yu} \Pi - \Gamma
		\end{array}\!\!\right]\right)$  into $m_{\xi}$ row blocks, with each block having $m_{x}$ rows. Denote the $i$th partitioned matrix and vector respectively by $\Upsilon_{\alpha,i}$ and $t_{1,0,i}$ with $i=1,2,\cdots, m_{\xi}$, counting from the ceiling. Define a constant matrix $T_{10}(\Gamma)$ as
\begin{displaymath}
\hspace*{-1.0cm}    T_{10}(\Gamma) \!=\! [t_{1,0,1} \;\; t_{1,0,2}\;\; \cdots \;\; t_{1,0,m_{\xi}}]
\end{displaymath}
Furthermore, define a constant vector $\gamma(\Gamma)$, as well as constant matrices $W_{t,0}$ and $W_{s,0}$, respectively as
{\small
\begin{eqnarray*}
& & \hspace*{-0.7cm}  \gamma(\Gamma) \!=\! {\rm\bf vec}\left((I-P_{0}D_{zv})\left[\!\!\begin{array}{c}
				B_{xv}\\
				D_{yv}
        \end{array}\!\! \right]^{\dag} \!
  \left[\!\!\begin{array}{c }  -B_{xu} \Pi  \\
				\Gamma \!-\! D_{yu} \Pi
		\end{array}\!\!\right] \!-\! P_{0}D_{zu}\Pi \right) -  \\
& & \hspace*{-0.0cm}
   \left\{\! I \!\otimes\! (P_{0}C_{zx}) \!+\! \left(\! I \!\otimes\! \left[ \!(P_{0}D_{zv} \!-\! I) \!
        \left[\!\!\!\begin{array}{c}
				B_{xv}\\
				D_{yv}
        \end{array}\!\!\! \right]^{\dag}\right]\right)
        \left(\! \left(\Xi^{T} \!\!\otimes\! I \right) \Upsilon_{t} - \begin{array}{c} \; \\ \; \end{array} \right. \right. \\
& & \hspace*{1.25cm} \left.\left. I \otimes
  \left[\!\!\begin{array}{c} A_{xx} \\
				C_{yx} \\
		\end{array}\!\!\right] \right) \right\} \Upsilon_{s}^{\dag}
  {\rm\bf vec}\left(\left[\!\!\begin{array}{c}
				B_{xv}\\
				D_{yv}
        \end{array}\!\! \right]_{l}^{\perp} \!
  \left[\!\!\begin{array}{c }  -B_{xu} \Pi  \\
				D_{yu} \Pi \!-\! \Gamma
		\end{array}\!\!\right]  \right)  \\
& & \hspace*{-0.7cm} W_{t,0} \!=\! \left\{I \!\otimes\! (P_{0}C_{zx}) \!+\!
   \left(\! I \!\otimes\! \left[(P_{0}D_{zv} \!-\! I) \!
        \left[\!\!\!\begin{array}{c}
				B_{xv}\\
				D_{yv}
        \end{array}\!\!\! \right]^{\dag}\right]\right)
        \left(\! \left(\Xi^{T} \!\!\otimes\! I \right) \Upsilon_{t} - \begin{array}{c} \; \\ \; \end{array} \right. \right.\\
& & \hspace*{4.6cm} \left.\left. I \otimes \!
  \left[\!\!\begin{array}{c} A_{xx} \\
				C_{yx} \\
		\end{array}\!\!\right] \right) \right\} \Upsilon_{s,r}^{\perp} \\
& & \hspace*{-0.7cm} W_{s,0} = I \otimes \left[(P_{0}D_{zv} - I)
        \left[\begin{array}{c}
				B_{xv}\\
				D_{yv}
        \end{array} \right]_{r}^{\perp}\right]
\end{eqnarray*}}
In addition, for each $i=1, 2, \cdots, m_{\theta}$, define a constant vector $w_{i}(\Gamma)$, as well as constant matrices $W_{t,i}$ and $W_{s,i}$, respectively as
{\small \begin{eqnarray*}
& & \hspace*{-0.7cm}  w_{i}(\Gamma) \!=\! {\rm\bf vec}\left\{P_{i}\left(D_{zu}\Pi \!+\! D_{zv} \left[\!\!\begin{array}{c}
				B_{xv}\\
				D_{yv}
        \end{array}\!\! \right]^{\dag} \!
        \left[\!\!\begin{array}{c }  -B_{xu} \Pi  \\
				\Gamma \!-\! D_{yu} \Pi
		\end{array}\!\!\right] \right) \right\}\!+\! \\
& & \hspace*{-0.4cm}
  \left( I\otimes(P_{i}C_{zx}) \!+\!
   \left[ I\otimes \left(P_{i}D_{zv}\!
        \left[\!\!\!\begin{array}{c}
				B_{xv}\\
				D_{yv}
        \end{array}\!\!\! \right]^{\dag}\right)\right] \times \right.  \\
& & \hspace*{-0.4cm}
        \left. \left[\! \left(\Xi^{T} \otimes I\right) \Upsilon_{t} - I \otimes \left[\!\!\begin{array}{c} A_{xx} \\
				C_{yx} \\
		\end{array}\!\!\right] \right] \right) \Upsilon_{s}^{\dag}
  {\rm\bf vec}\left(\left[\!\!\begin{array}{c}
				B_{xv}\\
				D_{yv}
        \end{array}\!\! \right]_{l}^{\perp} \!
  \left[\!\!\begin{array}{c }  -B_{xu} \Pi  \\
				D_{yu} \Pi  \!-\! \Gamma
		\end{array}\!\!\right]  \right)  \\
& & \hspace*{-0.7cm} W_{t,i} \!=\!\! \left(\!\! I \!\otimes\! (P_{i}C_{zx}) \!+\!\!
   \left[\! I \!\otimes\!\! \left(\! P_{i}D_{zv}\!
        \left[\!\!\!\begin{array}{c}
				B_{xv}\\
				D_{yv}
        \end{array}\!\!\! \right]^{\!\dag}\!\right) \!\!\right] \!\!\!\left[\! \left(\Xi^{T} \!\!\otimes\! I\right) \Upsilon_{t} \!-\! I \!\otimes\! \left[\!\!\!\begin{array}{c} A_{xx} \\
				C_{yx} \\
		\end{array}\!\!\!\right] \right]\!\!\right) \!\Upsilon_{s,r}^{\perp} \\
& & \hspace*{-0.7cm} W_{s,i} = I \otimes \left( P_{i}D_{zv}
        \left[\begin{array}{c}
				B_{xv}\\
				D_{yv}
        \end{array} \right]_{r}^{\perp}\right)
\end{eqnarray*}}
Clearly, all these matrices and vectors can be calculated from available information about System $\mathbf{\Sigma}_{p}$.

In the above definitions, the values of some vectors and matrices depend on the RTIM $\Gamma$.  To clarify this dependence, they are expressed as a vector/matrix valued function of this matrix. This clarification is helpful in the following discussions, which are about robustness of the suggested recovery procedure against estimation errors in an estimate for the RTIM $\Gamma$.

Let $\alpha_{t}$, $\alpha_{s}$, $\alpha_{t,i}|_{i=1}^{m_{\theta}}$ and $\alpha_{s,i}|_{i=1}^{m_{\theta}}$ be some vector variables having appropriate dimensions. Denote the composite vector variable $\left[	{\alpha}^{T}_{t} \;  {\alpha}^{T}_{s} \; {\alpha}^{T}_{t,i}|_{i=1}^{m_{\theta}} \;
{\alpha}^{T}_{s,i}|_{i=1}^{m_{\theta}} \right]^{T}$ by ${\alpha}$. Define a vector valued function $e(\theta,\alpha,\Gamma)$, as well as two matrix valued functions $T_{1}(\alpha,\Gamma)$ and $R(\theta,\alpha)$, respectively as follows,
\begin{eqnarray*}
& & \hspace*{-1.0cm} e(\theta,\alpha,\Gamma) \!=\! \sum_{i=1}^{m_{\theta}} \! w_{i}(\Gamma)\theta_{i} \!+\!
\sum_{i=1}^{m_{\theta}} \! W_{t,i}\alpha_{t,i} \!+\! \sum_{i=1}^{m_{\theta}} \! W_{s,i}\alpha_{s,i} \!+ \\
& & \hspace*{4cm}
 W_{t,0}\alpha_{t} \!+\! W_{s,0}\alpha_{s} \!- \gamma(\Gamma) \\
& & \hspace*{-1.0cm}
T_{1}(\alpha,\Gamma) =T_{10}(\Gamma) \!+\!
\left[\!\!\begin{array}{cccc }  \Upsilon_{\alpha,1}\alpha_{t} & \Upsilon_{\alpha,2}\alpha_{t} & \cdots & \Upsilon_{\alpha,m_{\xi}}\alpha_{t}
		\end{array}\!\!\right]    \\
& & \hspace*{-1.0cm}
R(\theta,\alpha) = \left[\begin{array}{ccccc}
				1          & \theta_{1}    & \theta_{2}    &  \cdots & \theta_{m_{\theta}}    \\
                \alpha_{t} & \alpha_{t,1}  & \alpha_{t,2}  &  \cdots & \alpha_{t,m_{\theta}}  \\
                \alpha_{s} & \alpha_{s,1}  & \alpha_{s,2}  &  \cdots & \alpha_{s,m_{\theta}}
        \end{array} \right]
\end{eqnarray*}

With these definitions and symbols, a necessary and sufficient condition can be stated as the following theorem for Equation (\ref{eqn:pr-1}) to have a solution, while its proof is deferred to Appendix I.

\begin{Theorem}\label{theo:3}
There is a vector $\theta$ satisfying Equation (\ref{eqn:pr-1}) with a FCR $T_{1}$, if and only if the following rank constrained linear vector equation of the vector variables $\theta$ and $\alpha$, has a solution,
\begin{equation}
e(\theta,\alpha,\Gamma) =0, \;\; subject \;\; to\;
\left\{\begin{array}{l} \hspace*{-0.15cm} {\rm\bf rank}\! \left\{T_{1}(\alpha,\Gamma) \right\} \!=\! m_{\xi} \\
\hspace*{-0.15cm} {\rm\bf rank}\left\{R(\theta,\alpha) \right\} = 1 \end{array}\right.
\label{eqn:theo:3-1}
\end{equation}
\end{Theorem}

The above theorem reveals that recovering the value of the parameter vector $\theta$ of System $\mathbf{\Sigma}_{p}$ from its RTIM $\Gamma$, is equivalent to finding a solution to a linear equation, that is, Equation (\ref{eqn:theo:3-1}), under some rank restraints on two matrices that depend on the variables affinely. This kind of problems has found wide applications in image process, system identification, model reduction, etc., and various efficient methods have been suggested for solving them \cite{rfp2010, tbls2024}.

From the proof of Theorem \ref{theo:3}, it is clear that the vector valued function $e(\theta,\alpha,\Gamma)$ is actually an error vector for Equation (\ref{eqn:pr-1}).

In actual parameter recovery, a RTIM $\Gamma$ may be computed from a high order model which is well adopted in model reduction \cite{abg2020,Astolfi2010}, or estimated from experimental data that is the case of system identification \cite{Zhou2025}. In both cases, errors are unavoidable for this matrix due to computation accuracy restrictions or measurement errors and process disturbances, etc.

When only an estimate, say $\widehat{\Gamma}$, is available for the RTIM $\Gamma$ of System $\mathbf{\Sigma}_{p}$ at the matrix pair $(\Xi,\;\Pi)$, a natural way to recover the value of its parameter vector $\theta$ is to perform the following constrained minimization problem, with respect to the vector variables $\overline{\theta}$ and $\overline{\alpha}$,
\begin{equation}
\hspace*{-0.0cm}
   \min_{\overline{\theta},\: \overline{\alpha}}\left|\left| e(\overline{\theta},\overline{\alpha},\widehat{\Gamma}) \right|\right|\;
subject \; to\;
\left\{\begin{array}{l} \hspace*{-0.15cm} \overline{\theta} \!\in\! \mathbf{\Theta},\; {\rm\bf rank}\left\{R(\overline{\theta},\overline{\alpha}) \right\} \!=\! 1    \\
\hspace*{-0.15cm} {\rm\bf rank}\! \left\{T_{1}(\overline{\alpha},\widehat{\Gamma}) \right\} \!=\! m_{\xi} \end{array}\right.
\label{eqn:pr-28}
\end{equation}
in which the vector variable $\overline{\theta}$ is constituted from some scalar variables $\overline{\theta}_{i}|_{i=1}^{m_{\theta}}$, while the vector variable $\overline{\alpha}$ from some vector variables $\overline{\alpha}_{t}$, $\overline{\alpha}_{s}$, $\overline{\alpha}_{t,i}|_{i=1}^{m_{\theta}}$ and $\overline{\alpha}_{s,i}|_{i=1}^{m_{\theta}}$, completely in the same way as that for the vectors ${\theta}$ and ${\alpha}$.

In this minimization, the vector norm $||\star||$ can be an arbitrary one that has some application significance. A widely adopted one is the Euclidean norm \cite{adm2020,mpr2014,zyl2018}. In this case, minimization of the cost function of Equation (\ref{eqn:pr-28}) is equivalent to that of $e^{T}\!(\overline{\theta},  \overline{\alpha}, \widehat{\Gamma})\,e(\overline{\theta},  \overline{\alpha}, \widehat{\Gamma})$, which is well known as a least squares estimation, and is immediate from the definition of the Euclidean norm of a real vector.

Recall that the FCR property of a matrix is generic. It can therefore be declared that when the matrices $A(\theta)$ and $\Xi$ do not have a common eigenvalue, any matrix that is close to the matrix $T_{1}$ of Equation (\ref{eqn:pr-1}) is almost surely of FCR, meaning that in this case, the rank constraint ${\rm\bf rank}\! \left\{T_{1}(\overline{\alpha},\widehat{\Gamma}) \right\} \!=\! m_{\xi}$ can be neglected in the aforementioned minimization.

In many applications, direct influences do not exist from a system input to a system output, meaning that in Equation (\ref{plant-3}), $D_{\#\S}$ with $\#=y$ or $z$, and $\S=u$ or $v$, is usually a zero matrix. In addition, measurements are often directly performed on some state variables that have clear significance in physics, chemistry, biology, etc., meaning that the matrix $C_{yx}$ of Equation (\ref{plant-3}) can usually be assumed to take the form $[I_{m_{y}}\;\; 0_{m_{y}\times(m_{x}-m_{y})}]$. In this case, if $ C_{zx} [0_{(m_{x}-m_{y})\times m_{y}}\;\; I_{m_{x}-m_{y}}]^{T} = 0$ is further satisfied, then the aforementioned parameter recovery procedure from a RTIM $\Gamma$ can be reduced to a least squares estimation without any rank constraints. The details are omitted for their obviousness from Equation (\ref{eqn:pr-2}) and close similarities to the proof of Theorem \ref{theo:3}, as well as for space considerations.

Taking account of the fact that the vector variable $\overline{\alpha}$ is usually of a high dimension, the nuclear norm based optimization method \cite{rfp2010,tbls2024} is adopted in this paper for obtaining an estimate for the parameter vector $\theta$, in which the rank constraint on the matrix $T_{1}(\overline{\alpha},\widehat{\Gamma})$ is neglected. However, rather than directly penalizing the difference between the nuclear norm and maximum singular value of the matrix $R(\overline{\theta},\overline{\alpha})$, which is widely adopted for handling the rank 1 constraint of a matrix \cite{rfp2010,tbls2024}, an instrumental matrix $R_{in}$ that has the same dimension as the matrix $R(\overline{\theta},\overline{\alpha})$ is introduced, which is required to satisfy the rank 1 condition. More precisely, rather than the rank constrained minimization of Equation (\ref{eqn:pr-28}), an unconstrained minimization is performed with respect to vector variables $\overline{\theta}$ and $\overline{\alpha}$, as well as an instrumental matrix variable $R_{in}$, in which the following cost function $J(\overline{\theta},\: \overline{\alpha}, R_{in})$ is adopted,
\begin{eqnarray}
& & \hspace*{-1.80cm}
   J(\overline{\theta},\overline{\alpha}, R_{in}) \!=\! \frac{\left|\left| e(\overline{\theta},\overline{\alpha},\widehat{\Gamma}) \right|\right|_{2}^{2}}{2}  \!+\! \nonumber\\
& & \hspace*{0.45cm}
   \lambda_{1} \!
   \left[\! \frac{\left|\left| R(\overline{\theta},\overline{\alpha}) \! - \! R_{in}\right|\right|_{F}^{2}}{2} \!\!+\! \lambda_{2} \! \! \sum_{i=2}^{m_{\theta} \!+\! 1} \!\!\sigma_{i}( R_{in}) \!\right]
\label{eqn:ns-3}
\end{eqnarray}
Here, $\lambda_{1}$ and $\lambda_{2}$ are some penalty factors that take positive values and can be tuned in actual applications.

Note that $\sum_{i=2}^{m_{\theta}+1} \!\!\sigma_{i}( R_{in})=||R_{in}||_{\star} - ||R_{in}||_{2}$ can be established directly from definitions of the associated matrix norms, meaning that the 3rd term of the cost function $J(\overline{\theta},\: \overline{\alpha}, R_{in})$ is associated with the difference of two convex functions. This association makes the minimization of this cost function mathematically hard. To overcome these difficulties, an iterative gradient descent procedure is adopted. In each iteration, $||R_{in}||_{2}$ is linearized at the value of the matrix $R(\overline{\theta},\overline{\alpha})$ of the same iteration.

Specifically, the following algorithm is adopted in this paper.

\renewcommand{\labelenumi}{\rm\bf S\arabic{enumi})}
\renewcommand{\labelenumii}{\rm\bf s\arabic{enumii})}
\setcounter{enumi}{0}
\addtocounter{enumi}{-1}
\vspace{-0.10cm}
\begin{enumerate}
\setlength{\topsep}{-0.3cm}
\setlength{\itemsep}{-0.0cm}
\item Set an initial value for the vectors $\overline{\theta}$ and $\overline{\alpha}$, as well as the matrix $R_{in}$, respectively as $\overline{\theta}^{[0]}$, $\overline{\alpha}^{[0]}$ and $R_{in}^{[0]}$. Set also an iteration control number $\varepsilon_{it}$, and a step length $\delta_{it}$. Assign the iteration number $k$ as $k=0$.

\item At the triple $(\overline{\theta}^{[k\!-\!1]},\overline{\alpha}^{[k\!-\!1]}, R_{in}^{[k\!-\!1]})$, calculate the partial differential of the cost function $J(\overline{\theta},\overline{\alpha}, R_{in})$ respectively with respect to $\overline{\theta}$ and $\overline{\alpha}$. Denote them by ${\partial J(\overline{\theta}^{[k\!-\!1]},\overline{\alpha}^{[k\!-\!1]}, R_{in}^{[k\!-\!1]})}/{\partial \overline{\theta}}$ and ${\partial J(\overline{\theta}^{[k\!-\!1]},\overline{\alpha}^{[k\!-\!1]}, R_{in}^{[k\!-\!1]})}/{\partial \overline{\alpha}}$, respectively. Update $\overline{\theta}^{[k]}$ and $\overline{\alpha}^{[k]}$ as follows, and denote the updated  $R(\overline{\theta}^{[k]},\overline{\alpha}^{[k]})$ by $R^{[k]}$ for brevity.

\begin{eqnarray*}
& & \hspace{-0.1cm}
\overline{\theta}^{[k]} \!=\! \overline{\theta}^{[k\!-\!1]} \!- \frac{\partial J(\overline{\theta}^{[k\!-\!1]},\overline{\alpha}^{[k\!-\!1]}, R_{in}^{[k\!-\!1]})}{\partial \overline{\theta}}\delta_{it} \\
& & \hspace{-0.1cm}
\overline{\alpha}^{[k]} \!=\! \overline{\alpha}^{[k\!-\!1]} \!- \frac{\partial J(\overline{\theta}^{[k\!-\!1]},\overline{\alpha}^{[k\!-\!1]}, R_{in}^{[k\!-\!1]})}{\partial \overline{\alpha}}\delta_{it}
\end{eqnarray*}

\item Let $R^{[k]} \!=\! \sum_{i=1}^{m_{\theta}+1} \!\!\sigma_{i}( R^{[k]})u_{i}^{[k]}v_{i}^{[k],T}$ denote an SVD for the matrix $R^{[k]}$, in which $\sigma_{i}( R^{[k]})$, $u_{i}^{[k]}$ and $v_{i}^{[k]}$ are respectively its $i$th singular values, and the associated left and right singular vectors. Minimize the following cost function $\overline{J}^{[k]}(R_{in})$ with respect to $R_{in}$. Denote the optimal value by $R_{in}^{[k]}$.
\begin{displaymath}
\hspace*{-0.0cm}
\overline{J}^{[k]}(R_{in}) \!=\! \left|\left| R^{[k]} \!-\! R_{in}\right|\right|_{F}^{2} \!+ \lambda_{2} \!  \left(||R_{in}||_{\star} \!-\! u_{1}^{[k],T}R_{in}v_{1}^{[k]} \right)
\end{displaymath}

\item Increase the iteration number $k$ by one. Repeat Steps S2) and S3) until $\left|J(\overline{\theta}^{[k]},\overline{\alpha}^{[k]},R_{in}^{[k]}) \!-\! J(\overline{\theta}^{[k\!-\!1]},\overline{\alpha}^{[k\!-\!1]},R_{in}^{[k\!-\!1]})\right| \!\leq\! \varepsilon_{it}$.

\end{enumerate}
\vspace{-0.25cm}

Note that $u_{1}^{[k],T}R^{[k]}v_{1}^{[k]} \!=\! ||R^{[k]}||_{2}$. The term $u_{1}^{[k],T}\!\!R_{in}v_{1}^{[k]}$ in the above optimization procedure is actually a linearization of $||R_{in}||_{2}$ at $R^{[k]}$. This approach is widely adopted in minimizing a concave function, and known as a concave-convex procedure \cite{rfp2010,tbls2024,yr2003}. On the other hand, the cost function $\overline{J}^{[k]}(R_{in})$ is strictly convex, and its optimum has an analytic expression. These properties are quite attractive, and are one of the major motivations for introducing the instrumental matrix variable $R_{in}$.\footnote[1]{Define a cost function $\widehat{J}^{[k]}(\overline{\theta},\overline{\alpha})$ as $\widehat{J}^{[k]}(\overline{\theta},\overline{\alpha}) \!=\! \frac{1}{2} || e(\overline{\theta},\overline{\alpha},\widehat{\Gamma}) ||_{2}^{2}  \!+\! \frac{\lambda_{1}}{2} \! || R(\overline{\theta},\overline{\alpha}) \! - \! R_{in}^{[k-1]}||_{F}^{2}$. Then, update for $\overline{\theta}^{[k]}$ and $\overline{\alpha}^{[k]}$ can in principle also be performed through minimizing this cost function, noting that it is strictly convex for any positive $\lambda_{1}$. In addition, from the linear dependence of $e(\overline{\theta},\overline{\alpha},\widehat{\Gamma})$ and $R(\overline{\theta},\overline{\alpha})$ on $\overline{\theta}$ and $\overline{\alpha}$, it is not very difficult to understand that this minimization is actually a standard least squares data-fitting problem, and an analytic formula for the optimal values can be obtained without significant difficulties. However, as the vector $\overline{\alpha}$ usually has a high dimension and matrix inversions are required in the associated computations, numerical stability issues may arise in this update, unless these inversions have an analytic expression.}

More precisely, the following conclusions can be established for the optimization of the cost function $\overline{J}^{[k]}(R_{in})$, while their proof is deferred to Appendix I.

\begin{Theorem}\label{theo:5}
Assume that there exists an integer $\tau$ in the set $\{0,1,\cdots, m_{\theta}\}$, such that $\sigma_{\tau +1}( R^{[k]}) < \lambda_{2}$. Then the matrix $R_{in}^{[k]}$ defined as follows is the unique global optimizer of the cost function $\overline{J}^{[k]}(R_{in})$,
\begin{equation}
\hspace*{-0.1cm}
R_{in}^{[k]} \!=\! \sigma_{1}(R^{[k]})u_{1}^{[k]}v_{1}^{[k],T} \!\!+\! \sum_{i=2}^{\tau}\!\left( \!\sigma_{i}(R^{[k]}) \!-\! \lambda_{2}\right)u_{i}^{[k]}v_{i}^{[k],T}
\label{eqn:theo:5-1}
\end{equation}
\end{Theorem}

From this theorem, it is clear that if the penalty factor $\lambda_{2}$ is appropriately selected, such that $\sigma_{1}(R^{[k]}) \gg \sigma_{2}(R^{[k]})-\lambda_{2}$, then the matrix $R_{in}^{[k]}$ is very close to a rank one matrix. This is also an attractive property for the associated optimizations, noting that the rank of the matrix $R(\overline{\theta},\overline{\alpha})$ is required to be equal to 1.

Compared with the parametric estimation algorithm of \cite{Zhou2025}, the above parameter recovery procedure does not put any assumptions on system matrices of System $\mathbf{\Sigma}_{p}$, and is therefore applicable to a much larger class of systems.

However, it is worthwhile to emphasize that although our simulations show that the suggested recovery algorithm has a good convergence record, and is not very sensitive to initial value selections, there are still no theoretical guarantees for these important properties. Further efforts are necessary for settling this essential issue.

\section{Robustness of the Parameter Recovery Procedure}

When an exact RTIM $\Gamma$ is used in the above minimizations, Theorem \ref{theo:3} makes it clear that the optimal value of the cost function  in both Equations (\ref{eqn:pr-28}) and (\ref{eqn:ns-3}) is equal to zero, which is reached by the actual value of the parameter vector $\theta$. But when an estimate $\widehat{\Gamma}$ for the RTIM $\Gamma$ is used, this minimum value is generally greater than zero, and its value depends on magnitude of errors in this estimate. In addition, the optimal value of the cost function is usually achieved by a vector that is different from the actual value of the parameter vector $\theta$.

As actual value is rarely available for a RTIM $\Gamma$ of System $\mathbf{\Sigma}_{p}$, it is important to assure that the suggested method for recovering the parameter vector $\theta$ is not very sensitive to errors in a given estimate for this RTIM $\Gamma$, meaning that a small error in its estimate should not result in a significant error in the recovered value for the parameter vector $\theta$. Obviously, a necessary condition for a recovery method to be robust against errors in an estimated RTIM is that the parameter vector $\theta$ is recoverable.

To discuss this robustness issue, define matrices $R_{xv}(\theta)$, $R_{xx}(\theta)$, $R_{yv}(\theta)$ and $R_{yx}(\theta)$ respectively as follows,
\begin{eqnarray*}
& & \hspace*{-1.20cm}
R_{xv}(\theta) \!=\!
                \left\{ \left(\left[ C_{zx} T_{1}(\theta) \!+\!  D_{zu} \Pi \right]^{T} \!\!\left[ I \!-\! D_{zv} P(\theta) \right]^{-\!T} \!\right)
         \otimes \right. \\
& & \hspace*{3.50cm} \left.\left(B_{xv} \!\left[ I \!-\! P(\theta)D_{zv} \right]^{-\!1} \right) \right\} \!\Psi \\
& & \hspace*{-1.20cm}
R_{xx}(\theta) \!=\!
        I \otimes \left(B_{xv}  \left[ I \!-\! P(\theta)D_{zv} \right]^{-1} \!\!P(\theta) C_{zx}  \!+\!
        A_{xx}\right) \!-\! \Xi^{T} \!\otimes\! I
         \\
& & \hspace*{-1.20cm}
R_{yv}(\theta) \!=\!
                \left\{\left(\left[ C_{zx} T_{1}(\theta) \!+\!  D_{zu} \Pi \right]^{T}\!\!\left[  I \!-\! D_{zv} P(\theta)  \right]^{-\!T}\right)
         \otimes \right. \\
& & \hspace*{3.50cm}        \left. \left( D_{yv} \left[ I \!-\! P(\theta)D_{zv} \right]^{-\!1} \right)\right\} \!\Psi  \nonumber\\
& & \hspace*{-1.20cm}
R_{yx}(\theta) \!=\!
        I \otimes \left(D_{yv}  \left[ I \!-\! P(\theta)D_{zv} \right]^{-1} \!\!P(\theta) C_{zx}  \!+\!
        C_{yx}\right)
\end{eqnarray*}
Here, the parameter vector $\theta$ is explicitly expressed for both the matrix $T_{1}$ of Theorem \ref{theo:2} and the newly defined matrices, clarifying dependence of these matrices on its actual value.

With these symbols, the following conclusions can be obtained for the robustness of the suggested parameter recovery method. Their proof can be found in Appendix I.

\begin{Theorem}\label{theo:4}
Assume that the parameter vector $\theta$ of System $\mathbf{\Sigma}_{p}$ is globally recoverable from its RTIM $\Gamma$, and $A(\theta)$ does not have any eigenvalue $\lambda$ that makes $\lambda I-\Xi$ singular. Then the suggested recovery procedure for the parameter vector $\theta$ is robust against errors in an estimate for the RTIM $\Gamma$, if and only if
the matrix $\left[ R_{yv}(\theta)\;\;  R_{yx}(\theta) \right] \left[ R_{xv}(\theta)\;\;   R_{xx}(\theta)\right]_{r}^{\perp}$ is of FCR.
\end{Theorem}

Recall that when the matrices $A(\theta)$ and $\Xi$ do not have an equal eigenvalue, each element of the matrix $T_{1}(\theta)$ that satisfies Equation (\ref{eqn:pr-1}) is a rational function of the parameter vector $\theta$. From the definition of the matrix $P(\theta)$, as well as the adopted assumptions of Theorem \ref{theo:4}, it is clear that each element of the aforementioned four matrices, that is, $R_{xv}(\theta)$, $R_{xx}(\theta)$, $R_{yv}(\theta)$ and $R_{yx}(\theta)$, is also a rational function of this parameter vector, provided that the adopted assumptions are satisfied simultaneously. Then similar arguments as those for Corollaries \ref{coro:1} and \ref{coro:2} show that although the condition of Theorem \ref{theo:4} depends on the actual value of the parameter vector $\theta$, its verification can be performed in a pragmatic way, through some random samples of the set $\mathbf{\Theta}$ that contains all of its possible values.

From Equation (\ref{eqn:pr-26}), it is clear that when the conditions of Theorem \ref{theo:4} are satisfied, and estimation errors for the RTIM $\Gamma$ are sufficiently small, then the Euclidean norm of errors in the parameter recovery, that is, $||\delta_{\theta}||_{2} $, is upper bounded by $|| [ I_{m_{\theta}}\; 0] [ R_{xv}(\theta)\;   R_{xx}(\theta)]_{r}^{\perp} \! ([ R_{yv}(\theta)\;  R_{yx}(\theta) ] [ R_{xv}(\theta)\;   R_{xx}(\theta)]_{r}^{\perp})^{\dag} ||_{2}||\delta_{\Gamma}||_{2}$, giving a quantitative measure for robustness of the suggested parameter recovery procedure against RTIM estimation errors.

Note that the matrices $R_{xv}(\theta)$, $R_{xx}(\theta)$, $R_{yv}(\theta)$ and $T_{1}(\theta)$ depend also on the matrix pair $(\Xi,\;\Pi)$. Equation (\ref{eqn:pr-26}) in the proof of Theorem \ref{theo:4} also reveals that an optimal selection for this matrix pair that leads to the minimal magnitude of parameter recovery errors, depends on the actual estimation errors for the RTIM $\Gamma$. This is in general impossible in actual applications. Numerical simulations in the next section, however, show that for a robust recovery procedure, a selection is possible for parameters of this matrix that are close to their optimal values.

\section{Numerical Example}

To illustrate characteristics of the suggested recovery procedure, as well as effectiveness in recovery accuracy improvement through  introducing a derivative of the TFM of a system, some numerical simulations are performed in this section, in which the natural frequency and damping ration of a single-input single-output LTI plant are directly recovered from information about its tangential interpolations. The transfer function of this plant is assumed to have the following structure,
\begin{displaymath}
H(s,\theta)= k\frac{(s+r_{z})(s^{2}+2\zeta_{z}\omega_{z}s+\omega_{z}^{2})}{(s+r_{p,1})(s+r_{p,2})(s^{2}+2\zeta_{p}\omega_{p}s+\omega_{p}^{2})}
\end{displaymath}
in which the values of $k$, $r_{z}$, $\zeta_{z}$, $\omega_{z}$, $r_{p,1}$ and $r_{p,2}$ are assumed known, while the values of $\zeta_{p}$ and $\omega_{p}$ are to be recovered. That is, $\theta=[\zeta_{p}\;\; \omega_{p}]^{T}$.

Rather than a 2nd order system and a more complicated one, the above plant is selected mainly for satisfying the requirement $m_{x}\geq m_{\xi}$, and clearly showing differences in parameter recoveries that are due to a derivative introduction. Moreover, the following values are chosen for the plant parameters,
\begin{eqnarray*}
& & \hspace*{-1.3cm}
k=6.0000,\hspace{0.15cm} r_{z}=2.0000,\hspace{0.15cm} \zeta_{z}=2.0000\times 10^{-1},\hspace{0.15cm} \omega_{z}=8.0000  \\
& & \hspace*{-1.3cm}
r_{p,1}\!=\! 3.0000,\hspace{0.1cm}r_{p,2} \!=\! 5.0000,\hspace{0.1cm}\zeta_{p} \!=\! 1.0000 \!\times\! 10^{-1},\hspace{0.1cm}\omega_{p} \!=\! 5.0000
\end{eqnarray*}

Note that $r_{p,2}=\omega_{p}$ and $\omega_{z}$ is not very far from $\omega_{p}$. From the frequency-magnitude characteristics of this plant, which is shown in Figure \ref{fig:1}, this parameter value choice is believed capable of making the recovery of $\zeta_{p}$ and $\omega_{p}$ more difficult.

Some straightforward algebraic manipulations give the following system matrices of a state space model for this plant,
\begin{eqnarray*}
& & \hspace*{-1.25cm} A(\theta)=\left[\!\! \begin{array}{cccc}
-r_{p,1} & r_{p,1}-r_{z} & 0                & 0 \\
0      & -r_{p,2}        & 0                & 0 \\
0      & 0             & 0                & 1 \\
-1     & 1             & -\omega_{p}^{2}  & -2\zeta_{p}\omega_{p} \end{array} \!\!  \right], \hspace{0.25cm}
B(\theta)=\left[\!\! \begin{array}{c}
0 \\
k \\
0 \\
0 \end{array} \!\!  \right] \\
& & \hspace*{-1.25cm}
C(\theta)=\left[\!\! \begin{array}{cccc}
-1 & 1 & \omega_{z}^{2}-\omega_{p}^{2} &
2(\zeta_{z}\omega_{z} -\zeta_{p}\omega_{p}) \end{array} \!\!  \right], \hspace{0.25cm}
D(\theta)=0
\end{eqnarray*}

\renewcommand{\thefigure}{\arabic{figure}}
\setcounter{figure}{0}
\vspace{-0.0cm}
\begin{figure}[!ht]
\vspace{-0.0cm}
\begin{center}
\includegraphics[width=3.0in]{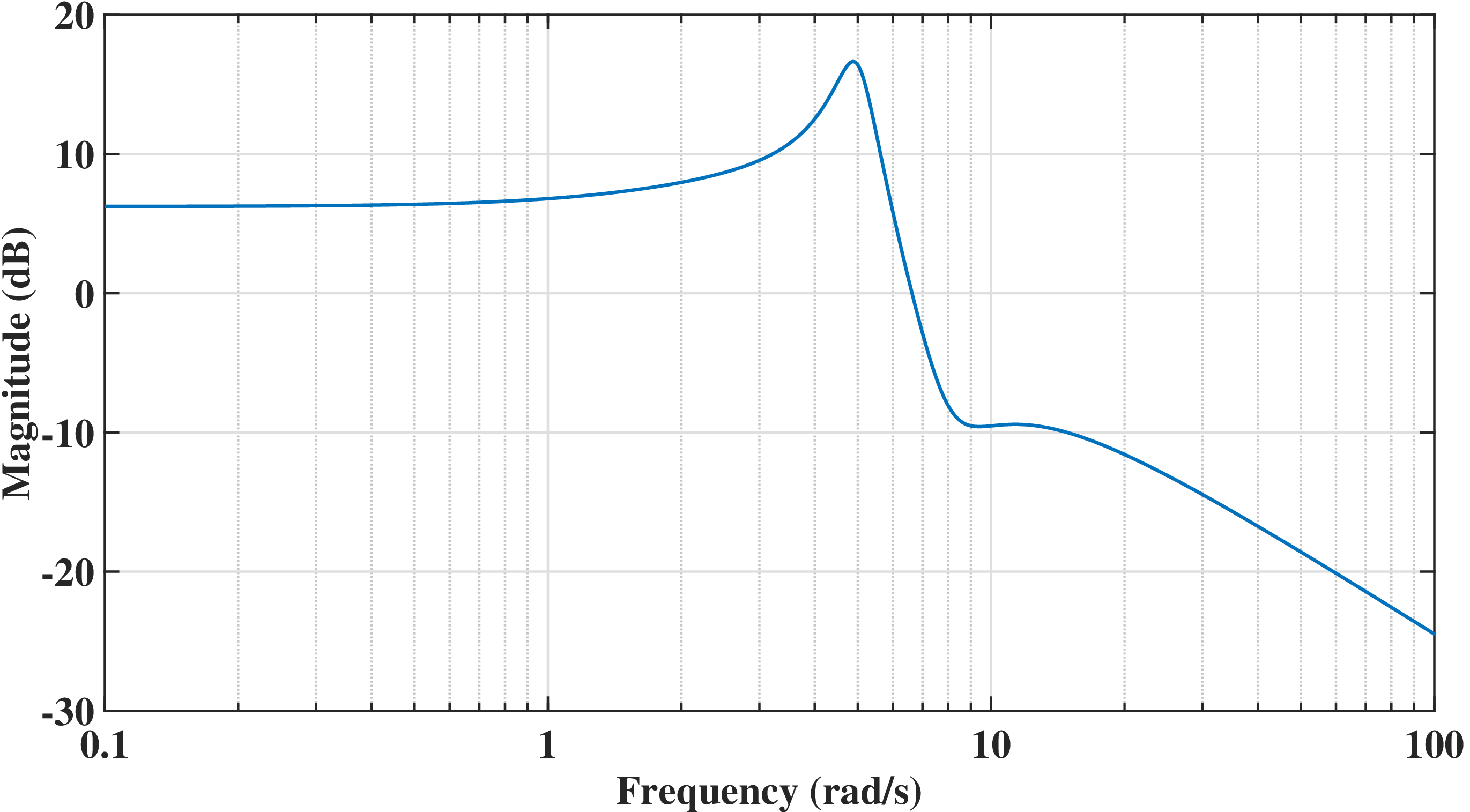}
\vspace{-0.6cm}\hspace*{5cm} \caption{Frequency-magnitude characteristics of the plant.}
\label{fig:1}
\vspace{-0.2cm}
\end{center}
\end{figure}

Define the matrix $P(\theta)$ as
\begin{displaymath}
P(\theta)=\left[\!\! \begin{array}{ccc}
0 & \omega_{p} & 0       \\
\zeta_{p}     & 0             & \omega_{p} \end{array} \!\!  \right]
\end{displaymath}
Then these system matrices can be expressed by the LFT form of Equations (\ref{plant-3}) and (\ref{plant-4}). The associated system matrices are given in Appendix II.

To investigate effectiveness of derivatives of a system TFM in improving parameter recovery accuracy, the following two sets are selected in simulations for the matrices $\Xi$ and $\Pi$,
\begin{eqnarray*}
& & \hspace*{-1.25cm} \Xi^{[1]}=\left[\!\! \begin{array}{cccc}
\sigma_{s}^{[1]} & \omega_{s}^{[1]} & 1                & 0 \\
-\omega_{s}^{[1]}     & \sigma_{s}^{[1]}        & 0                & 1 \\
0      & 0             & \sigma_{s}^{[1]}                & \omega_{s}^{[1]} \\
0     & 0             & -\omega_{s}^{[1]}  & \sigma_{s}^{[1]} \end{array} \!\!  \right], \hspace{0.25cm}
\Pi^{[1]} =\left[\!\! \begin{array}{c}
1 \\
1 \\
0 \\
0 \end{array} \!\!  \right]^{T}  \\
& & \hspace*{-1.25cm}  \Xi^{[0]}=\left[\!\! \begin{array}{cccc}
\sigma_{s,1}^{[0]} & \omega_{s,1}^{[0]} & 0                & 0 \\
-\omega_{s,1}^{[0]}      &  \sigma_{s,1}^{[0]}       & 0                & 0 \\
0      & 0             & \sigma_{s,2}^{[0]}                & \omega_{s,2}^{[0]} \\
0     & 0             & -\omega_{s,2}^{[0]}  & \sigma_{s,2}^{[0]} \end{array} \!\!  \right], \hspace{0.25cm}
\Pi^{[0]} =\left[\!\! \begin{array}{c}
1 \\
1 \\
1 \\
1 \end{array} \!\!  \right]^{T}
\end{eqnarray*}

Clearly from Theorem 1 of \cite{Zhou2025}, when the matrix pair $(\Xi^{[1]}, \Pi^{[1]})$ is adopted, information about both the values and the first order derivatives of $H(s)$ at $\sigma_{s}^{[1]} \pm j\omega_{s}^{[1]}$ is included in the associated RTIM, denote it by $\Gamma^{[1]}$. On the other hand, when the matrix pair $(\Xi^{[0]}, \Pi^{[0]})$ is adopted, the associated RTIM, denote it by $\Gamma^{[0]}$, contains information only about the values of $H(s)$ at $\sigma_{s,i}^{[0]} \pm j\omega_{s,i}^{[0]}$ with $i=1,2$. To clarify these differences, the superscripts $[0]$ and  $[1]$ are used to indicate the inclusion of the zeroth and first order derivative. On the other hand, from the dimensions of the associated matrices, both $\Gamma^{[1]}$ and $\Gamma^{[0]}$ belong to $\mathbb{R}^{1\times 4}$, revealing an equal data length in parameter recovery. In the following discussions, their $k$th column elements are respectively denoted by $\gamma^{[1]}_{k}$ and $\gamma^{[0]}_{k}$, $1\leq k \leq 4$. In addition, for each $i=0,1$, an $\Gamma^{[i]}$ associated estimate for $\theta$ is denoted by $\widehat{\theta}^{[i]}$, with its elements respectively by $\widehat{\zeta}_{p}^{[i]}$ and $\widehat{\omega}_{p}^{[i]}$.

In this simulation, $\lambda_{1}=2.0000$ and $\lambda_{2}=10.0000$ are selected for the penalty factors in the cost function of Equation (\ref{eqn:ns-3}). Note that with the selected plant parameter values, the magnitude of the plant frequency response, that is, $H(j\omega,\theta)$, reaches its maximum approximately at $\omega=4.8816rad/s$, which can also be understood from Figure \ref{fig:1}. Note also that $2\zeta_{p}\omega_{p}=1.0000$. The following procedure is adopted for selecting parameters of the matrix $\Xi^{[i]}$, $i=1,2$, with the purpose of finding their values that are close to the best which lead to the most accurate parameter recovery.\footnote[2]{While $\sigma_{s}^{[1]}$, $\sigma_{s,1}^{[0]}$ and $\sigma_{s,2}^{[0]}$ can also be chosen by a random search, it is not performed, as the selected parameters have already clearly reflected properties of the suggested recovery procedure. On the other hand, our simulations show that a $\sigma_{s}^{[1]}=\sigma_{s,1}^{[0]}=\sigma_{s,2}^{[0]}$ closer to $-\zeta_{p}\omega_{p}$ usually increases recovery accuracy for the parameter vector $\theta$.} 

\begin{itemize}
\item Assign $\sigma_{s}^{[1]}=\sigma_{s,1}^{[0]}=\sigma_{s,2}^{[0]}=-5.0000\times 10^{-2}$.
\item Randomly generate 4 independent numbers according to a normal distribution with its expectation and standard deviation respectively being $0.0000$ and $1.7000\times 10^{-2}$. Denote them by $\varepsilon_{i}|_{i=1}^{4}$.
\item Randomly generate 100 independent $\omega_{s}^{[1]}$s according to a continuous uniform distribution over the interval $[4.0000,\; 6.0000]$. For each generated random sample, calculate the $\sigma_{s}^{[1]} \pm j\omega_{s}^{[1]}$ associated RTIM $\Gamma^{[1]}$, and generate its estimate as $\widehat{\Gamma}^{[1]} ={\rm\bf row}\left\{\gamma^{[1]}_{i}(1+\varepsilon_{i})|^4_{i=1}\right\}$. Recover  $\theta$ with this $\widehat{\Gamma}^{[1]}$ and measure the magnitude of estimation errors by
    $\sqrt{(\zeta_{p}-\widehat{\zeta}_{p}^{[1]})^{2}/\zeta_{p}^{2} + (\omega_{p}-\widehat{\omega}_{p}^{[1]})^{2}/\omega_{p}^{2}}$. Among all the generated $\omega_{s}^{[1]}$s, select the $\omega_{s}^{[1]}$ that leads to the minimal magnitude of estimation errors as the desirable value.
\item Randomly generate 100 independent $(\omega_{s,1}^{[0]}, \omega_{s,2}^{[0]})$ pairs according to a continuous uniform distribution over the area $[4.0000,\; 6.0000]\times [4.0000,\; 6.0000]$. Perform $(\omega_{s,1}^{[0]}, \omega_{s,2}^{[0]})$ selection similarly as those in selecting $\omega_{s}^{[1]}$.
\end{itemize}

Using this random search, the following values are selected for the parameters of the matrix $\Xi^{[i]}$, $i=1,2$,
\begin{displaymath}
\omega_{s,1}^{[0]} \!=\! 4.4179rad/s, \hspace{0.15cm} \omega_{s,2}^{[0]} \!=\! 4.5306rad/s, \hspace{0.15cm} \omega_{s}^{[1]} \!=\! 4.4799rad/s
\end{displaymath}

It is interesting to note that $\omega_{s}^{[1]}\approx 0.5(\omega_{s,1}^{[0]}+\omega_{s,2}^{[0]})$ and $|\omega_{s,1}^{[0]}-\omega_{s,2}^{[0]}|$ is small. This may possibly imply that information about derivatives of a TFM is really important in parameter recovery, recalling that derivative of a function is the limit of the ratio between differences in its value and variable.

With these selected parameters, the following computations are performed to investigate properties of the suggested recovery procedure, as well as effectiveness of introducing derivative information in parameter recovery.

\begin{itemize}
\item Generate the RTIMs ${\Gamma}^{[0]}$ and ${\Gamma}^{[1]}$ of System $\mathbf{\Sigma}_{p}$ respectively at the matrix pairs $(\Xi^{[0]},\,\Pi^{[0]})$ and $(\Xi^{[1]},\,\Pi^{[1]})$.
\begin{itemize}
\item Generate a random vector $\varepsilon \!=\! {\rm\bf row}\{\varepsilon_{i}|^4_{i=1}\}$, in which $\varepsilon_{i}$, $1 \!\leq\! i \!\leq\! 4$, are independent of each other and obey a normal distribution with its expectation and standard deviation respectively being $0.0000$ and $1.7000\times 10^{-\!1}\!\!$.
\item Generate estimates for the RTIMs $\Gamma^{[0]}$ and $\Gamma^{[1]}$ respectively as $\widehat{\Gamma}^{[i]} ={\rm\bf row}\{\gamma^{[i]}_{k}(1+\varepsilon_{k})|^4_{k=1}\}$, $i=0,1$.
\item Compute estimates for $\theta$, that is, $\widehat{\theta}^{[0]}$ and $\widehat{\theta}^{[1]}$, respectively with the RTIM estimates $\widehat{\Gamma}^{[0]}$ and $\widehat{\Gamma}^{[1]}$, through performing the suggested recovery algorithm.
\item Record the magnitude of relative recovery error for each parameter, that is, $100\times|(\zeta_{p}\!-\!\widehat{\zeta}_{p}^{[i]})/\zeta_{p}|$ and $100\times|(\omega_{p}\!-\!\widehat{\omega}_{p}^{[i]})/\omega_{p}|$, $i=0,1$, as well as their ratios $r_{\zeta}$ and $r_{\omega}$, that are calculated respectively as $r_{\zeta} \!=\!|(\zeta_{p}\!-\!\widehat{\zeta}_{p}^{[1]})/(\zeta_{p}\!-\!\widehat{\zeta}_{p}^{[0]})|$ and $r_{\omega} \!=\! |(\omega_{p}\!-\!\widehat{\omega}_{p}^{[1]})/(\omega_{p}\!-\!\widehat{\omega}_{p}^{[0]})|$.
\end{itemize}
\item Repeat the above calculations 300 times.
\item Rearrange the recorded magnitudes of relative estimation errors, as well as their ratios, according to the Euclidean norm of the generated random vector $\varepsilon$.
\end{itemize}

Obviously, a $r_{\zeta}$ smaller than $1.0000$ means that the estimate $\widehat{\zeta}_{p}^{[1]}$ is better than the estimate $\widehat{\zeta}_{p}^{[0]}$, and the converse is also true. These statements remain valid for $r_{\omega}$. Hence, $r_{\zeta}$ and $r_{\omega}$ can be regarded as a measure for effectiveness of derivative information introduction in parameter recovery.

\vspace{-0.0cm}
\begin{figure}[!ht]
\vspace{-0.2cm}
\begin{center}
\includegraphics[width=3.0in]{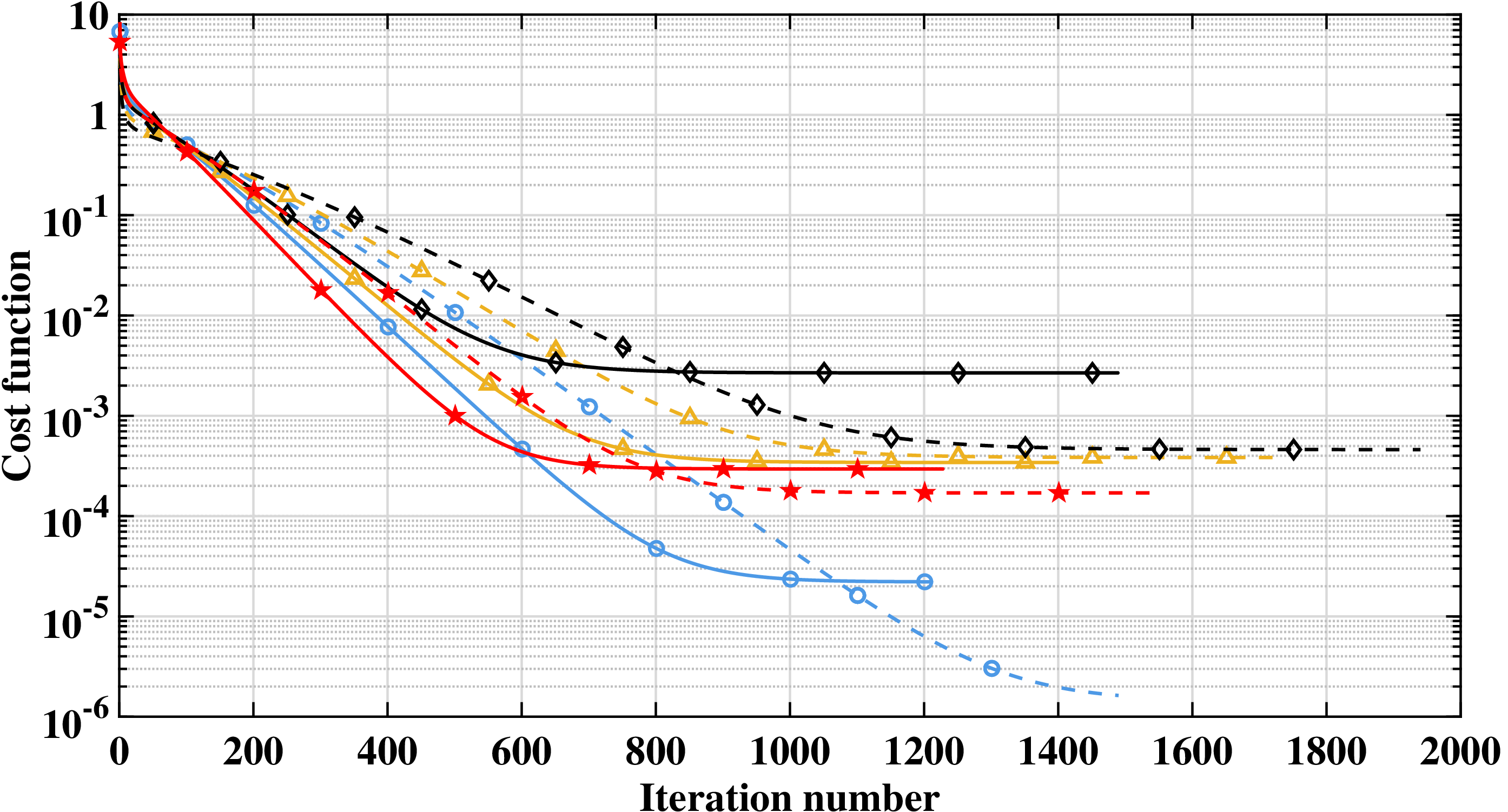}
\vspace{-0.6cm}\hspace*{5cm} \caption{Typical cost function curves in parameter recovery.}
{\small \hspace*{-2.45cm} $-\:-$: $\widehat{\Gamma}^{[0]}$ based recovery; $-\!-$: $\widehat{\Gamma}^{[1]}$ based recovery. \\
 \hspace*{-0.2cm}$\circ$: $\varepsilon \!=\! [1.9477\!\times\! 10^{-\!2}\;\; 7.9221\!\times\! 10^{-\!3}\;\; 4.4740 \!\times\! 10^{-\!3}\;\; -\! 3.6647 \!\times\! 10^{-\!2}]$; \\
$\triangle$: $\varepsilon \!=\! [2.5656\!\times\! 10^{-\!2}\;\: -\!4.8425\!\times\! 10^{-\!2}\;\: -\!1.6932\!\times\! 10^{-\!1}\;\: 1.5780\!\times\! 10^{-\!4}]$; \\
$\diamondsuit$: $\varepsilon \!=\! [-\!2.7062\!\times\! 10^{-\!1}\; -\!4.2597\!\times\! 10^{-\!2}\; -\!4.7609\!\times\! 10^{-\!2}\; 1.2500\!\times\! 10^{-\!1}]$; \\
\hspace*{-0.2cm}$\ast$: $\varepsilon \!=\! [8.7601\!\times\! 10^{-\!2}\; -\!2.8810\!\times\! 10^{-\!1}\; 1.8755\!\times\! 10^{-\!1}\; 5.4091\!\times\! 10^{-\!2}]$.}
\label{fig:2}
\vspace{-0.2cm}
\end{center}
\end{figure}

It is worthwhile to mention that in the above simulations, the standard deviation for each element of the random vector $\varepsilon$ is selected $10$ times as large as that associated with the parameter selection for the matrices $\Xi^{[0]}$ and $\Xi^{[1]}$, aiming at a confirmation for the robustness of the suggested recovery procedure.

In this numerical example, the matrix $\Upsilon_{s}$ of Equation (\ref{eqn:pr-29}) happens to be of FCR. On the basis of Equation (\ref{eqn:pr-8}), this means that the vector variable $\alpha_{t}$, and therefore the vector variables $\alpha_{t,1}$ and $\alpha_{t,2}$, are no longer available in the composite vector variable $\alpha$.

\begin{figure}[!ht]
\vspace{-0.0cm}
\begin{center}
\includegraphics[width=3.0in]{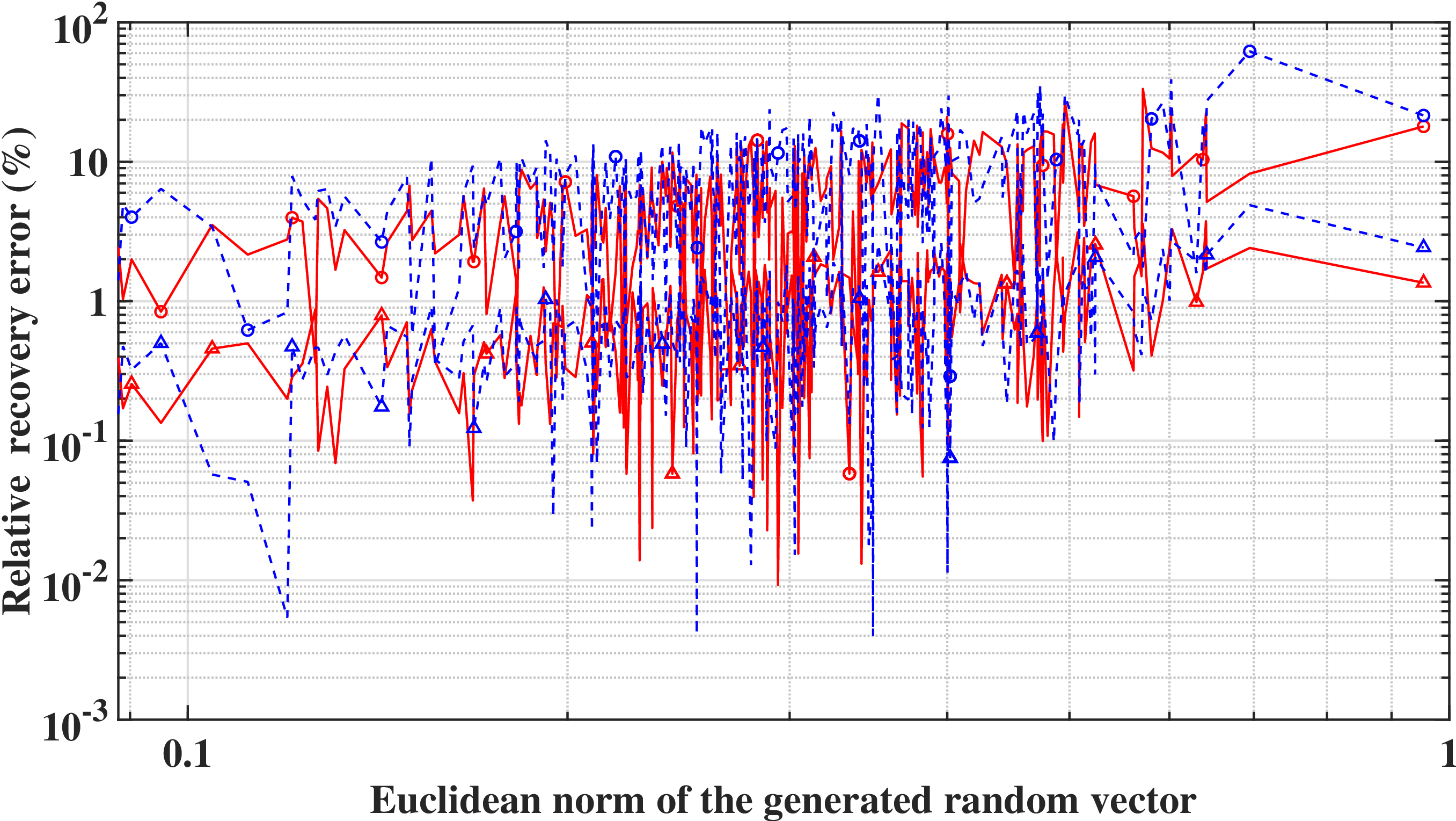}

(a) relative error. $-\:-$: $\widehat{\Gamma}^{[0]}$ based recovery; $-\!-$: $\widehat{\Gamma}^{[1]}$ based recovery. $\circ$: damping ratio $\zeta_{p}$; $\triangle$: natural frequency $\omega_{p}$.

\vspace{0.30cm}
\includegraphics[width=3.0in]{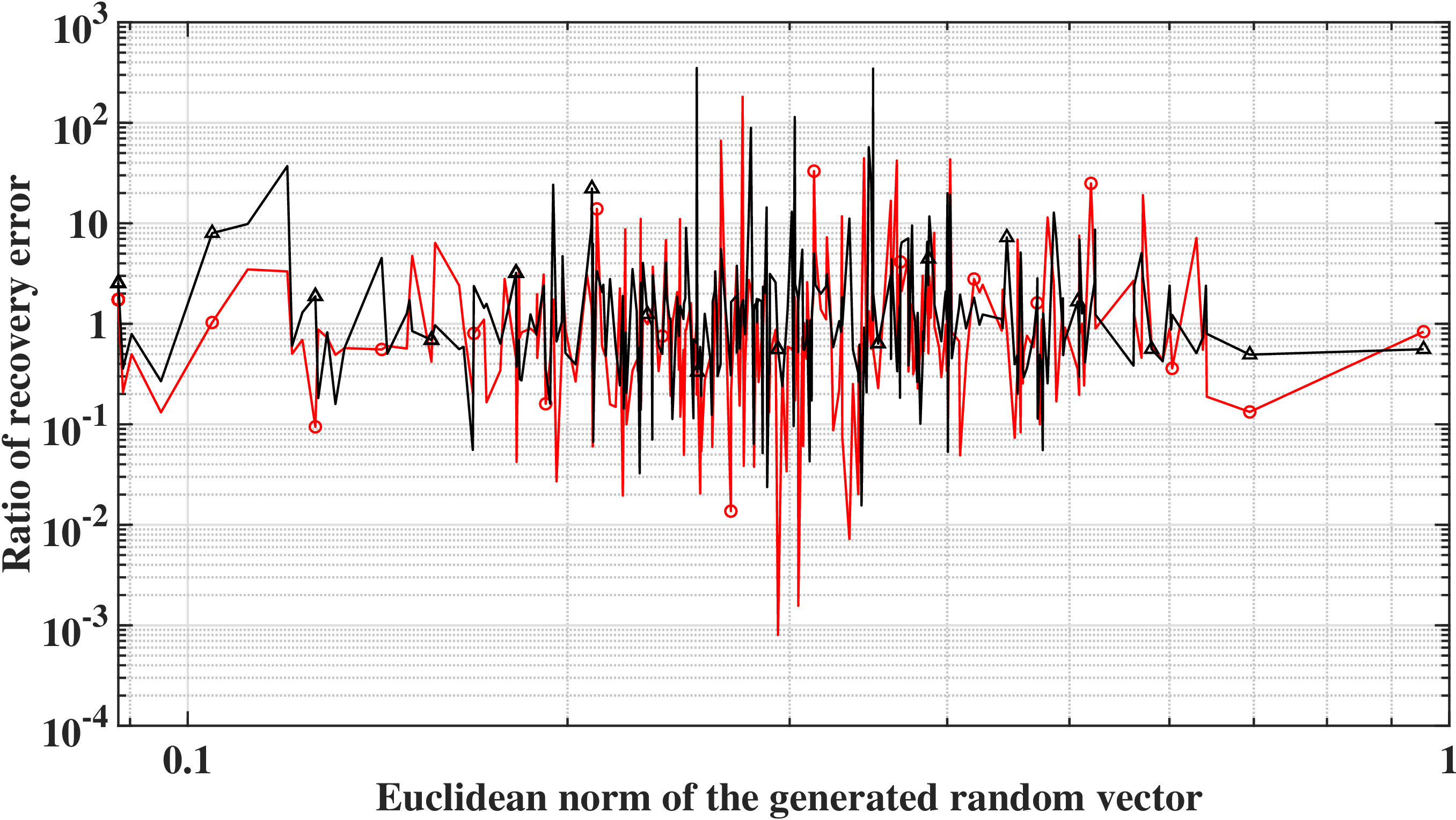}

(b) error ratio. $\circ$: $\zeta_{p}$ associated $r_{\zeta}$; $\triangle$: $\omega_{p}$ associated $r_{\omega}$.

\vspace{-0.65cm}\hspace*{5cm} \caption{Relative parameter recovery errors and their ratio.}
\label{fig:3}
\vspace{-0.4cm}
\end{center}
\end{figure}

In parameter recovery, the iteration control number and the step length are selected respectively as $\varepsilon_{it}=1.0000\times 10^{-10}$ and $\delta_{it}=0.0500$, while the maximum iteration number is set to be $2500$. In addition, the initial values for parameter estimates are selected as $\overline{\zeta}_{p}^{[0]} \!=\! 1.0000$ and $\overline{\omega}_{p}^{[0]} \!=\! 10.0000$. Moreover, the initial value for $\overline{\alpha}_{s}$, that is, $\overline{\alpha}_{s}^{[0]}$, is obtained through an unconstrained least squares estimation, that minimizes  $||e(\overline{\theta}^{[0]}, \overline{\alpha},\widehat{\Gamma})||_{2}^{2}$ and has an analytic global minimizer, in which $\overline{\alpha}_{s,1}$ and $\overline{\alpha}_{s,2}$ are respectively set as $\overline{\alpha}_{s,1}=\overline{\zeta}_{p}^{[0]}\overline{\alpha}_{s}$ and $\overline{\alpha}_{s,2}=\overline{\omega}_{p}^{[0]}\overline{\alpha}_{s}$. Furthermore, the initial value for $R_{in}$  is set as $R_{in}^{[0]}=R(\overline{\theta}^{[0]},\overline{\alpha}^{[0]})$. Here, $\overline{\theta}^{[0]}=[\overline{\zeta}_{p}^{[0]}\;\; \overline{\omega}_{p}^{[0]}]^{T}$ and $\overline{\alpha}^{[0]}=[\overline{\alpha}_{s}^{[0],T}\;\;
\overline{\zeta}_{p}^{[0]}\overline{\alpha}_{s}^{[0],T}\;\;   \overline{\omega}_{p}^{[0]}\overline{\alpha}_{s}^{[0],T}]^{T}$.

Clearly, these initial values are not very close to their actual ones, which are intentionally chosen to see robustness of the suggested recovery algorithm against initial value selections.

All the iterative computations converge, although some of them are significantly faster than the others. Some typical curves of the cost function are given in Figure \ref{fig:2}. In all cases, the cost function decreases monotonically with the iteration number, while the $\widehat{\Gamma}^{[1]}$ based iterations generally converge faster than those based on $\widehat{\Gamma}^{[0]}$. This figure also reveals that significant differences exist among the optimal values of the cost function with a different value of the random vector $\varepsilon$, reflecting that the optimal parameter values of the matrices $\Xi^{[0]}$ and $\Xi^{[1]}$ depend also on the disturbance strength, as well as that the randomly selected parameter values for these two matrices may not be very close to their optimal ones in some of the parameter recoveries.

Figure \ref{fig:3} gives relative errors of each parameter recovery with the generated random samples for the RTIM, as well as their ratios $r_{\zeta}$ and $r_{\omega}$, arranged according to the Euclidean norm of $\varepsilon$. Clearly, the natural frequency ${\omega}_{p}$ is in general recovered with a higher accuracy than the damping ratio ${\zeta}_{p}$. Moreover, estimate based on information about both the value and the derivative of the plant transfer function is in most cases more accurate than that only on its value,\footnote[3]{The total sample number associated with $r_{\omega}<1.0000$ is slightly smaller than $150$, that is, half of the total number of the generated samples. This may be considered due to computation errors and parameter selection for the matrix $\Xi^{[i]}$, $i=1,2$, noting that most of the relative errors for $\omega_{p}$ recovery are less than $1.0000\%$, and optimal $\Xi^{[0]}$ and $\Xi^{[1]}$  depend also on disturbance strength.} especially in recovering the damping ratio ${\zeta}_{p}$ and when the disturbances on the RTIM have a low strength. The former is not out of prediction, as damping ratios are widely believed hard to estimate, especially when they take a small value. The latter is an issue that has not been well observed yet, limited to the best of our knowledge.

To clarify relations among parameter recovery accuracy, strength of disturbances on RTIM, and effectiveness of different kinds of plant information, the Euclidean norms of the randomly generated disturbance vectors are divided into five intervals. For each of these intervals, the number of the random samples belonging to it is counted, as well as those respectively satisfying $r_{\zeta}<1.0000$ and $r_{\omega}<1.0000$. The results are reported in Table 1, showing clear advantages of introducing derivative information in parameter recovery.

\section{Concluding Remarks}

In this paper, parameter recovery is attacked for an LTI system, using information about its TFM value and derivative, in which system matrices are assumed to be some LFTs of the parameters. A necessary and sufficient condition is obtained for parameter recoverability, which reduces to a FCR requirement on a constant matrix in some particular cases. A method is suggested for parameter recovery, and its robustness against data imperfectness has been established. Numerical simulations provide a good convergence record for the suggested method, and also show that it is not sensitive to initial value selections. These simulations also reveal that derivative information about a system TFM may be quite helpful in its parameter recovery.

 \vspace{-0.0cm}
\begin{table}[t]
\vspace*{-0.00cm}
\caption{\vspace*{-0.00cm} Distribution of the generated samples  }
\begin{center}
\vspace{-0.0cm}
\begin{tabular}{|c||c|c|c|c|c|c|c|}
\hline
\hspace*{-0.18cm} ${\small ||\varepsilon||_{2}}$ \hspace{-0.28cm} & \hspace*{-0.18cm}{\small [0.05,0.20)} \hspace{-0.28cm} & \hspace*{-0.18cm}{\small [0.20,0.35)} \hspace{-0.28cm} & \hspace*{-0.18cm}{\small [0.35,0.50)} \hspace{-0.28cm} & \hspace*{-0.18cm}{\small [0.50,0.65)} \hspace{-0.28cm} & \hspace*{-0.18cm}{\small [0.65,1.00)} \hspace{-0.28cm} & \hspace*{-0.30cm}{\small Sum} \hspace{-0.40cm} \\
\hline
\hline
\hspace*{-0.18cm}${\small r_{\zeta}<1}$ \hspace{-0.28cm} & {\small 31} & {\small 97} & {\small 43} & {\small 14} & {\small 2} & \hspace*{-0.08cm}{\small 187}\hspace{-0.35cm}\\ \hline

\hspace*{-0.18cm}${\small r_{\omega}<1}$ \hspace{-0.28cm} & {\small 28} & {\small 70} & {\small 36} & {\small 10} & {\small 2} & \hspace*{-0.08cm}{\small 146}\hspace{-0.35cm}\\ \hline

\hspace*{-0.18cm}{\small Total} \hspace{-0.28cm} & {\small 46} & {\small 153} & {\small 75} & {\small 24} & {\small 2} &
\hspace*{-0.08cm}{\small 300}\hspace{-0.35cm}\\ \hline
\end{tabular}
\vspace{-0.5cm}
\end{center}
\label{table:1}
\vspace{-0.00cm}
\end{table}

Further works include method developments for efficiently obtaining tangential interpolations for a system outside the imaginary axis of the complex plane, as well as providing theoretical guarantees for the convergence of the proposed parameter recovery procedure, and advantages of a system TFM derivative in parameter recovery.

\renewcommand{\labelenumi}{\rm\bf A\arabic{enumi})}

\renewcommand{\theequation}{A\arabic{equation}}
\setcounter{equation}{0}

\vspace{-0.0cm}
\small
\section*{\small \hspace*{-0.0cm} Appendix I: Proof of Some Technical Results}

\addtolength{\abovedisplayskip}{-0.10cm}
\addtolength{\belowdisplayskip}{-0.10cm}

\addtolength{\abovedisplayshortskip}{-0.10cm}
\addtolength{\belowdisplayshortskip}{-0.10cm}

\vspace{-0.00cm}

\noindent\textbf{Proof of Theorem \ref{theo:1}:}
Assume that there is another value of the parameter vector $\theta$ that belongs to the set $\mathbf{\Theta}$, denote it by $\overline{\theta}$, such that the corresponding System $\mathbf{\Sigma}_{p}(\overline{\theta})$ has the same RTIM $\Gamma(\theta)$ at the matrix pair $(\Xi,\;\Pi)$ as that of System $\mathbf{\Sigma}_{p}(\theta)$\footnote[4]{Here, with a little abuse of terminology, differences have not been explicitly distinguished for System $\mathbf{\Sigma}_{p}(\theta)$ between the parameter vector $\theta$ and its value, in order to avoid awkward expressions. This treatment is also adopted in other statements and equations, provided that confusions can not be easily caused.}, that is, $\Gamma(\overline{\theta}) = \Gamma(\theta)$. Then according to the definition of the RTIM $\Gamma(\overline{\theta})$, it can be claimed that there exists a matrix $X(\overline{\theta})$, such that
\begin{equation}\label{tan-int-3}
\hspace*{-0.25cm} A(\overline{\theta})X(\overline{\theta}) + B(\overline{\theta})\Pi - X(\overline{\theta})\Xi = 0, \hspace{0.25cm} 	
\Gamma(\theta) = C(\overline{\theta})X(\overline{\theta}) + D(\overline{\theta})\Pi  	
\end{equation}

Comparing Equation (\ref{tan-int-3}) with Equations (\ref{tan-int-1}) and (\ref{tan-int-2}), the following relations are obtained straightforwardly,
\begin{eqnarray}
& & \hspace*{-1.20cm}
    \left[ C({\theta})X({\theta}) + D({\theta})\Pi \right] -
    \left[ C(\overline{\theta})X(\overline{\theta}) + D(\overline{\theta})\Pi \right] = 0
\label{eqn-app:1} \\
& & \hspace*{-1.20cm}
    \left[ A({\theta})X({\theta}) + B({\theta})\Pi - X({\theta})\Xi \right] -
    \left[ A(\overline{\theta})X(\overline{\theta}) + B(\overline{\theta})\Pi - X(\overline{\theta})\Xi \right] =0	
\label{eqn-app:2}
\end{eqnarray}

Through some direct algebraic manipulations, these relations can be equivalently expresses as
\begin{eqnarray}
& & \hspace*{-0.00cm} \left[\!\begin{array}{cc}
				A({\theta}) - A(\overline{\theta})  &  B({\theta}) - B(\overline{\theta}) \\
				C({\theta}) - C(\overline{\theta})  &  D({\theta}) - D(\overline{\theta})
        \end{array}\!\right]
\left[\!\begin{array}{c}
				X({\theta}) \\
				\Pi
        \end{array}\!\right]   \nonumber\\
& & \hspace*{-0.35cm} = \left[\!\begin{array}{c}
				A({\theta}) \left[ X(\overline{\theta}) - X({\theta}) \right] -  \left[ X(\overline{\theta}) - X({\theta}) \right] \Xi\\
				C({\theta}) \left[ X(\overline{\theta}) - X({\theta}) \right]
        \end{array}\!\right]
\label{eqn-app:3}
\end{eqnarray}

Note that
\begin{eqnarray*}
& & \hspace*{-0.00cm} \left[ I - P(\theta)D_{zv} \right]^{-1}P(\theta) - \left[ I - P(\overline{\theta})D_{zv} \right]^{-1}P(\overline{\theta})  \\
& & \hspace*{-0.35cm} = \left[ I - P(\theta)D_{zv} \right]^{-1}P(\theta) - P(\overline{\theta}) \left[ I - D_{zv} P(\overline{\theta}) \right]^{-1}  \\
& & \hspace*{-0.35cm} =\left[ I - P(\theta)D_{zv} \right]^{-1} \left[ P(\theta)-P(\overline{\theta}) \right] \left[ I - D_{zv} P(\overline{\theta}) \right]^{-1}
\end{eqnarray*}
Substitute Equation (\ref{plant-3}) into Equation (\ref{eqn-app:3}), the next equality is established,
\begin{eqnarray}
& & \hspace*{-0.60cm}
        \left[\!\!\!\!\begin{array}{c}
				B_{xv}\\
				D_{yv}
        \end{array}\!\!\!\!\right]\!\!
        \left[ I \!-\! P(\theta)D_{zv} \right]^{-1}\!\! \left[ P(\theta) \!-\! P(\overline{\theta}) \right] \!\! \left[ I \!-\! D_{zv} P(\overline{\theta}) \right]^{-\!1}\!\!
        \left[\!\!\! \begin{array}{cc}
				C_{zx} & \!\!\!\!\! D_{zu}
        \end{array} \!\!\!\right] \!\!\left[\!\!\!\!\begin{array}{c}
				X({\theta}) \\
				\Pi
        \end{array}\!\!\!\!\right] \nonumber\\
& & \hspace*{-0.85cm} = \left[\!\!\!\begin{array}{c}
				A({\theta}) \left[ X(\overline{\theta}) - X({\theta}) \right] - \left[ X(\overline{\theta}) - X({\theta}) \right] \Xi\\
				C({\theta}) \left[ X(\overline{\theta}) - X({\theta}) \right]
        \end{array}\!\!\!\right]
\label{eqn-app:4}
\end{eqnarray}

From the definition of the matrix $P(\theta)$ given in Equation (\ref{plant-4}), as well as the definition of the matrix $\Psi$ given in Lemma \ref{lemma:4}, it is obvious that
\begin{displaymath}
\hspace*{-0.25cm}
P(\theta)-P(\overline{\theta}) = \sum\limits_{i=1}^{m_{\theta}}\left( \theta_{i} - \overline{\theta}_{i} \right) P_{i}, \hspace{0.25cm}
{\rm\bf vec}\left(P(\theta)-P(\overline{\theta})\right) = \Psi (\theta -\overline{\theta})
\end{displaymath}
in which $\overline{\theta}_{i}$, $i=1,2,\cdots,m_{\theta}$, stands for the $i$th row element of the vector $\overline{\theta}$.
Then under the condition that the matrix $\Psi$ is of FCR, it can be seen without significant difficulties that $P(\theta)-P(\overline{\theta})=0$ if and only if $\overline{\theta} = \theta$.

Assume that there is a vector $\overline{\theta}$ satisfying Equation (\ref{eqn-app:4}) with its right side different from a zero matrix. Then it is necessary that $P(\theta)-P(\overline{\theta})$ is also different from a zero matrix, implying that $\overline{\theta} \neq \theta$ when the matrix $\Psi$ is of FCR. It can therefore be declared from Equation (\ref{eqn-app:4}) that under the condition that the matrix $\Psi$ is of FCR, the value of the parameter vector $\theta$ can be uniquely recovered from its RTIM $\Gamma(\theta)$, which is equivalent to that $\overline{\theta}=\theta$ is the only solution to Equation (\ref{eqn-app:4}), only if
\begin{equation}
\left[\!\begin{array}{c}
				A({\theta}) \left[ X(\overline{\theta}) - X({\theta}) \right] - \left[ X(\overline{\theta}) - X({\theta}) \right] \Xi\\
				C({\theta}) \left[ X(\overline{\theta}) - X({\theta}) \right]
        \end{array}\!\right] = 0
\label{eqn-app:5}
\end{equation}
On the other hand, from the assumption that the matrices $A({\theta})$ and $\Xi$ do not share any eigenvalues, it can be easily understood from matrix theories \cite{gv1989,hj1991} that the equality of Equation (\ref{eqn-app:5}) is equivalent to $X(\overline{\theta}) = X({\theta})$.

On the contrary, assume that $\overline{\theta} \neq \theta$ is a solution to the following equation,
\begin{eqnarray}
& & \hspace*{-1.00cm}
        \left[\!\!\!\!\begin{array}{c}
				B_{xv}\\
				D_{yv}
        \end{array}\!\!\!\!\right]
        \left[ I - P(\theta)D_{zv} \right]^{-1} \left[ P(\theta)-P(\overline{\theta}) \right] \times \nonumber\\
& & \hspace*{1.35cm}
        \left[ I - D_{zv} P(\overline{\theta}) \right]^{-1}
        \left[ C_{zx} X({\theta}) +  D_{zu} \Pi \right]  = 0
\label{eqn-app:10}
\end{eqnarray}
Then from Equations (\ref{eqn-app:4}) and (\ref{eqn-app:5}), as well as the assumption that the matrices $A({\theta})$ and $\Xi$ do not share any eigenvalues, it can be declared that this solution also satisfies $X(\overline{\theta}) = X({\theta})$.

The above arguments clarify that under the assumption that the matrix $\Psi$ is of FCR, the value of the parameter vector $\theta$ of System $\mathbf{\Sigma}_{p}$ is recoverable from its RTIM $\Gamma(\theta)$ at the matrix pair $(\Xi,\;\Pi)$, if and only if Equation (\ref{eqn-app:10}) has a unique solution $\overline{\theta} = {\theta}$.

For each $i=1,2,\cdots,m_{\theta}$, denote $\theta_{i} - \overline{\theta}_{i}$ by $\delta_{i}$. Moreover, denote $\theta - \overline{\theta}$ by $\delta$, and define matrices $\overline{P}(\delta)$ and $\widetilde{P}(\delta)$ respectively as
\begin{displaymath}
\hspace*{-0.25cm}
\overline{P}(\delta) = \sum\limits_{i=1}^{m_{\theta}} \delta_{i} P_{i}, \hspace{0.25cm}
\widetilde{P}(\delta) = \overline{P}(\delta)\left[ I-D_{zv} P(\theta)\right]^{-1}
\end{displaymath}
Then it is straightforward from Equation (\ref{plant-4}) that
\begin{equation}
P(\overline{\theta}) = P(\theta) - \overline{P}(\delta)
\label{eqn-app:6}
\end{equation}

Moreover, from these definitions and relations, the following equalities are also obtained
\begin{eqnarray}
& & \!\!\!\!
\left[ P(\theta)-P(\overline{\theta}) \right] \left[ I - D_{zv} P(\overline{\theta}) \right]^{-1} \nonumber\\
&=& \!\!\!\!
\overline{P}(\delta)\left[ I - D_{zv} P({\theta}) + D_{zv} \overline{P}(\delta) \right]^{-1}         \nonumber\\
&=& \!\!\!\!
\overline{P}(\delta)\left[ I - D_{zv} P({\theta}) \right]^{-1}\left\{ I + D_{zv} \overline{P}(\delta)\left[ I - D_{zv} P({\theta}) \right]^{-1} \right\}^{-1} \nonumber\\
&=& \!\!\!\! \widetilde{P}(\delta)\left[ I + D_{zv} \widetilde{P}(\delta) \right]^{-1}         \nonumber\\
&=& \!\!\!\! \left[ I + \widetilde{P}(\delta) D_{zv} \right]^{-1} \widetilde{P}(\delta)
\label{eqn-app:7}
\end{eqnarray}

Substitute the last equality of Equation (\ref{eqn-app:7}) into the left side of Equation (\ref{eqn-app:10}). It can be declared from Lemma \ref{lemma:3} that, if $\overline{\theta}$ is a solution to Equation (\ref{eqn-app:10}), then there exists a matrix $\Phi$ such that
\begin{equation}
\hspace*{-0.20cm}
\left[ I \!-\! P(\theta)D_{zv} \right]^{-1}\! \left[ I \!+\! \widetilde{P}(\delta) D_{zv} \right]^{-1} \!\widetilde{P}(\delta)
        \left[ C_{zx}X({\theta}) \!+\! D_{zu} \Pi \right]
\!=\! \left[\!\!\!\!\begin{array}{c}
				B_{xv}\\
				D_{yv}
        \end{array}\!\!\!\!\right]_{r}^{\perp}\!\!\Phi
\label{eqn-app:8}
\end{equation}
which is equivalent to
\begin{eqnarray}
\hspace*{-0.50cm}
& & \!\!\!\!\widetilde{P}(\delta)\left\{
        \left[ C_{zx}X({\theta}) + D_{zu} \Pi \right] - \left[ I - D_{zv}P(\theta) \right] D_{zv}\left[\!\!\!\!\begin{array}{c}
				B_{xv}\\
				D_{yv}
        \end{array}\!\!\!\!\right]_{r}^{\perp}\!\!\Phi \right\}  \nonumber\\
&=& \!\!\!\! \left[ I - P(\theta)D_{zv} \right]\left[\!\!\!\!\begin{array}{c}
				B_{xv}\\
				D_{yv}
        \end{array}\!\!\!\!\right]_{r}^{\perp}\!\!\Phi
\label{eqn-app:9}
\end{eqnarray}
Here, $\Phi$ is clearly a real matrix that has $m_{\xi}$ columns, while the number of its rows is equal to the dimension of the right null space of the matrix $\left[ B^{T}_{xv}\;\; D^{T}_{yv}\right]^{T}$. That is, the smaller this right null space is, the less the element number of the matrix $\Phi$ is.

Note that the matrix $I - P(\theta)D_{zv}$ is invertible, while the matrix $\left[\!\!\!\!\begin{array}{c} B_{xv} \\ D_{yv} \end{array}\!\!\!\!\right]_{r}^{\perp}$ is of FCR. It can therefore be declared that their product is also of FCR, implying that the right side of Equation (\ref{eqn-app:9}) is equal to a zero matrix, if and only if the matrix $\Phi$ is a zero matrix.  Note also that $\widetilde{P}(\delta) = 0$ if and only if $\overline{P}(\delta) = 0$, as well as that ${\rm\bf vec}\left( \overline{P}(\delta) \right) = \Psi (\theta -\overline{\theta})$ and the matrix $\Psi$ is of FCR, implying that $\widetilde{P}(\delta)$ is equal to a zero matrix, if and only if the vector $\delta$ is a zero vector. Hence, if there is a $\Phi \neq 0$ satisfying Equation (\ref{eqn-app:9}), then it is necessary that $\widetilde{P}(\delta) \neq 0$, and therefore $\delta \neq 0$, meaning that the value of the parameter vector $\theta$ can not be uniquely recovered from the RTIM $\Gamma(\theta)$.

Substitute the definition of the matrix $\widetilde{P}(\delta)$ into Equation (\ref{eqn-app:9}). The above arguments and some straightforward algebraic manipulations show that the value of the parameter vector $\theta$ can be uniquely determined from the RTIM $\Gamma(\theta)$, if and only if the following bilinear matrix equation
\begin{eqnarray}
& & \hspace*{-0.6cm}\left(\sum\limits_{i=1}^{m_{\theta}}\delta_{i}P_{i} \!\!\right)\!
\left\{\! \left[ I \!-\! D_{zv} P(\theta)\right]^{-\!1} \!\left[C_{zx} X(\theta) \!+\! D_{zu} \Pi\right] \!-\! D_{zv} \!
\left[\!\!\! \begin{array}{c}
				B_{xv}\\
				D_{yv}
        \end{array} \!\!\!\right]_{r}^{\perp}\!\! \!\Phi \!\!\right\} \nonumber \\
& &  \hspace*{-0.9cm}= \left[ I - P(\theta)D_{zv} \right]
\left[\!\!\!\begin{array}{c}
				B_{xv}\\
				D_{yv}
        \end{array}\!\!\!\right]_{r}^{\perp} \Phi
\label{eqn-app-19}
\end{eqnarray}
has a unique solution $\Phi=0$ and $\delta_{i}=0$, $i=1,2,\cdots,m_{\theta}$.

Vectorize both sides of the above equation, and recall the definition of the matrix $\Psi$ given in Lemma \ref{lemma:4}. Equation (\ref{eqn-app-19}) can be equivalently rewritten as
\begin{eqnarray}
& & \hspace*{-0.0cm}\!\!
  \left(\left\{\left[ I-D_{zv} P(\theta)\right]^{-1}\left[C_{zx} X(\theta) + D_{zu} \Pi\right]\right\}^{T}\otimes I\right) \Psi \delta -
\nonumber \\
& & \hspace*{2.0cm}\!\!
  \sum\limits_{i=1}^{m_{\theta}}\left\{\delta_{i} \left[I\otimes \left(P_{i}D_{zv} \left[\!\!\!\begin{array}{c}
				B_{xv}\\
				D_{yv}
        \end{array}\!\!\!\right]_{r}^{\perp}\right)\right] {\rm\bf vec}(\Phi) \right\}
\nonumber \\
& & \hspace*{-0.7cm} =
  \left( I \otimes \left\{\left[ I - P(\theta)D_{zv} \right]
  \left[\!\!\!\begin{array}{c}
				B_{xv}\\
				D_{yv}
        \end{array}\!\!\!\right]_{r}^{\perp}\right\}\right) {\rm\bf vec}(\Phi)
\label{eqn-app-20}
\end{eqnarray}

Denote ${\rm\bf vec}(\Phi)$ by $\phi$. Equation (\ref{eqn-app-20}) can be further equivalently rewritten as
\begin{eqnarray}
& & \hspace*{-1.3cm}\!\!
  \left\{\left(\left\{\left[ I-D_{zv} P(\theta)\right]^{-1}\left[C_{zx} X(\theta) + D_{zu} \Pi\right]\right\}^{T}\otimes I\right) \Psi  -
  \right. \nonumber \\
& & \hspace*{-1.3cm}\!\!
  \left.{\rm\bf row} \!\left(\! \left.
  \left[I \!\otimes\! \left(P_{i}D_{zv} \!\left[\!\!\!\!\begin{array}{c}
				B_{xv}\\
				D_{yv}
        \end{array}\!\!\!\!\right]_{r}^{\perp}\right)\right] \phi\right|_{i=1}^{m_{\theta}} \!\right)\right\}\delta  \nonumber \\
& & \hspace*{-1.65cm}
=\!
  \left( I \!\otimes\! \left\{\left[ I \!-\! P(\theta)D_{zv} \right]
  \left[\!\!\!\!\begin{array}{c}
				B_{xv}\\
				D_{yv}
        \end{array}\!\!\!\!\right]_{r}^{\perp} \!\right\} \!\right) \phi
\label{eqn-app-21}
\end{eqnarray}
According to Lemma \ref{lemma:3}, there is a nonzero $\delta$ satisfying this equation, if and only if there exists a $\phi \neq 0$, such that  \begin{eqnarray}
& & \hspace*{-1.3cm}\!\!
  \left\{\left(\left\{\left[ I-D_{zv} P(\theta)\right]^{-1}\left[C_{zx} X(\theta) + D_{zu} \Pi\right]\right\}^{T}\otimes I\right) \Psi  -
  \right. \nonumber \\
& & \hspace*{-1.15cm}\!\!
  \left.{\rm\bf row} \!\left(\! \left.
  \left[I \!\otimes\! \left(P_{i}D_{zv} \!\left[\!\!\!\!\begin{array}{c}
				B_{xv}\\
				D_{yv}
        \end{array}\!\!\!\!\right]_{r}^{\perp}\right)\right] \phi\right|_{i=1}^{m_{\theta}} \!\right)\right\}_{\!l}^{\!\perp} \!\!\times \nonumber \\
& & \hspace*{-1.15cm}
  \left( I \!\otimes\! \left\{\left[ I \!-\! P(\theta)D_{zv} \right]
  \left[\!\!\!\!\begin{array}{c}
				B_{xv}\\
				D_{yv}
        \end{array}\!\!\!\!\right]_{r}^{\perp} \!\right\} \!\right) \phi \!=\! 0
\label{eqn-app-22}
\end{eqnarray}
That is, to guarantee that there does not exist a $\overline{\theta} \in \mathbf{\Theta}$ that is different from $\theta$ and is consistent with the RTIM $\Gamma(\theta)$, it is necessary and sufficient that Equation (\ref{eqn-app-22}) does not have a nonzero solution. This completes the proof. \hspace{\fill}$\Diamond$

\noindent\textbf{Proof of Corollary \ref{coro:1}:}
From the definition of left and right null spaces of a matrix, it is clear that $\left[ B^{T}_{xv} \;\; D^{T}_{yv} \right]_{l}^{\perp} D^{T}_{zv} = 0$ if and only if $D_{zv} \left[\!\!\! \begin{array}{c} B_{xv}\\ D_{yv} \end{array} \!\!\!\right]_{r}^{\perp}=0$. Hence, when the condition that $\left[ B^{T}_{xv} \;\; D^{T}_{yv} \right]_{l}^{\perp} D^{T}_{zv} = 0$ is satisfied, Equation (\ref{eqn-app-19}) in the proof of Theorem \ref{theo:1} reduces to the following linear equation.
\begin{eqnarray}
\hspace*{-0.0cm} & & \!\!\!\! \left(\sum\limits_{i=1}^{m_{\theta}}\delta_{i}P_{i}\right)
\left[ I-D_{zv} P(\theta)\right]^{-1}\left[C_{zx} X(\theta) + D_{zu} \Pi\right] \nonumber \\
\hspace*{-0.0cm} &=& \!\!\!\! \left[ I - P(\theta)D_{zv} \right]
\left[\!\!\!\begin{array}{c}
				B_{xv}\\
				D_{yv}
        \end{array}\!\!\!\right]_{r}^{\perp} \Phi \nonumber \\
\hspace*{-0.0cm} &=& \!\!\!\!
\left[\!\!\!\begin{array}{c}
				B_{xv}\\
				D_{yv}
        \end{array}\!\!\!\right]_{r}^{\perp} \Phi
\label{eqn-app:15}
\end{eqnarray}

%
Vectorizing both sides of the above equation, the following equivalent equation is obtained from the definition of the matrix $\Psi$,
\begin{eqnarray}
& & \hspace*{-1.0cm}\!\!
\left(\left\{\left[ I-D_{zv} P(\theta)\right]^{-1}\left[C_{zx} X(\theta) + D_{zu} \Pi\right]\right\}^{T}\otimes I\right) \Psi \delta + \nonumber \\
& & \!\! \hspace*{2.5cm}
\left( I \otimes \left[\!\!\!\begin{array}{c}
				B_{xv}\\
				D_{yv}
        \end{array}\!\!\!\right]_{r}^{\perp}\right) {\rm\bf vec}(-\Phi) =0
\label{eqn-app:16}
\end{eqnarray}
which has a unique solution of $\delta=0$ and $\Phi=0$ , if and only if the next matrix is of FCR,
\begin{displaymath}
\hspace*{-0.4cm}\!\! \left[ \left(\left\{\left[ I-D_{zv} P(\theta)\right]^{-1}\left[C_{zx} X(\theta) + D_{zu} \Pi\right]\right\}^{T}\otimes I\right) \Psi \;\;\;  I \otimes \left[\!\!\!\begin{array}{c}
				B_{xv}\\
				D_{yv}
        \end{array}\!\!\!\right]_{r}^{\perp}\right]
\end{displaymath}
that is equivalent to that the matrix of Equation (\ref{eqn:coro:1-1}) is of FRR.

When the matrix $C_{zx} X(\theta) + D_{zu} \Pi$ is of FRR, it is clear from Equation (\ref{eqn-app:10}) that the value of the parameter vector $\theta$ is recoverable from the RTIM $\Gamma(\theta)$, if and only if the following equation has a unique solution $\overline{\theta} =
\theta$, noting that the matrix $I - D_{zv} P(\overline{\theta})$ is guaranteed to be invertible from Assumption \ref{assum:1}.
\begin{equation}
\hspace*{-1.00cm}
        \left[\!\!\!\!\begin{array}{c}
				B_{xv}\\
				D_{yv}
        \end{array}\!\!\!\!\right]
        \left[ I - P(\theta)D_{zv} \right]^{-1} \left[ P(\theta)-P(\overline{\theta}) \right]  = 0
\label{eqn-app:17}
\end{equation}

Vectorize both sides of this equation. Then the following equivalent equation can be directly obtained from the definitions of the matrices $P(\theta)$, $P(\overline{\theta})$ and $\Psi$.
\begin{equation}
 \hspace*{-1.2cm}
\left(I \otimes \left\{\left[\!\!\!\begin{array}{c}
				B_{xv}\\
				D_{yv}
        \end{array}\!\!\!\right]\left[ I - P(\theta)D_{zv} \right]^{-1}\right\}\right)\Psi (\theta - \overline{\theta}) = 0
\label{eqn-app:18}
\end{equation}
which has a unique solution $\overline{\theta} = \theta$, if and only if the right null space of the following matrix consists only of a zero vector, meaning that this matrix is of FCR,
\begin{equation}
 \hspace*{-1.2cm}
\left(I \otimes \left\{\left[\!\!\!\begin{array}{c}
				B_{xv}\\
				D_{yv}
        \end{array}\!\!\!\right]\left[ I - P(\theta)D_{zv} \right]^{-1}\right\}\right)\Psi
\label{eqn-app:18-x}
\end{equation}

Note that the matrix $\Psi$ is assumed to be FCR. It can be directly claimed from Lemma \ref{lemma:2} that the matrix of Equation (\ref{eqn-app:18-x}) is of FCR, if and only if the following matrix is of FCR,
\begin{equation}
 \hspace*{-1.2cm}
\left[\!\! \begin{array}{c}
I \otimes \left\{\left[ B^{T}_{xv}\;\; D^{T}_{yv} \right]^{T}\left[ I - P(\theta)D_{zv} \right]^{-1}\right\} \\
        \Psi_{l}^{\perp}  \end{array} \!\!\right]
\label{eqn-app:18-y}
\end{equation}

Recall that System $\mathbf{\Sigma}_{p}$ is assumed to be well-posed for each $\theta \in \mathbf{\Theta}$. It can be further declared from the following equality that the matrix of Equation (\ref{eqn-app:18-y}) is of FCR, if and only if the matrix of Equation (\ref{eqn:coro:1-2}) is of FCR,
\begin{eqnarray}
& & \!\!\!\!\left[\!\!\! \begin{array}{c}
I \!\otimes\! \left\{\left[ B^{T}_{xv}\; D^{T}_{yv} \right]^{T} \!\!\left[ I \!-\! P(\theta)D_{zv} \right]^{-\!1}\!\right\} \\
        \Psi_{l}^{\perp}  \end{array} \!\!\!\right] \nonumber \\
&=& \!\!\!\!
\left[\!\!\! \begin{array}{c} I \!\otimes\! \left[ B^{T}_{xv}\;\; D^{T}_{yv} \right]^{T}\\
        \Psi_{l}^{\perp}\left(I \!\otimes\! \left[ I \!-\! P(\theta)D_{zv} \right]\right)  \end{array} \!\!\!\right]
        \left(  I \!\otimes\!\left[ I \!-\! P(\theta)D_{zv} \right] \right)^{-\!1}
\label{eqn-app:18-z}
\end{eqnarray}
This completes the proof. \hspace{\fill}$\Diamond$

\noindent\textbf{Proof of Corollary \ref{coro:2}:}
Assume that there is a nonzero real vector $\phi$ such that the matrix of Equation (\ref{eqn:coro:2-1}) is of FRR. Then its left null space is trivial, that is, this null space only consists of a zero row vector. Hence the condition of Equation (\ref{eqn:theo:1}) is not satisfied by this nonzero $\phi$, meaning that the value of the parameter vector $\theta$ is not recoverable.

On the other hand, assume that the matrix of Equation (\ref{eqn:coro:2-2}) is not of FCR. From the well-posedness assumption about System $\mathbf{\Sigma}_{p}$, that is, Assumption \ref{assum:1}, as well as the following equality,
\begin{eqnarray}
& & \left[\!\! \begin{array}{c} \left[C_{zx} X(\theta) + D_{zu} \Pi\right]^{T} \otimes I \\
 \Psi_{l}^{\perp}\left(\left[ I-D_{zv} P(\theta)\right]^{T}\otimes I\right) \end{array} \!\! \right]
 \left(\left[ I-D_{zv} P(\theta)\right]\otimes I\right)^{-T} \nonumber \\
&=&
\left[\!\! \begin{array}{c} \left\{\left[C_{zx} X(\theta) + D_{zu} \Pi\right]^{T}\left[ I-D_{zv} P(\theta)\right]^{-T}\right\}\otimes I \\
 \Psi_{l}^{\perp} \end{array} \!\! \right]
\label{eqn:coro:2-3}
\end{eqnarray}
it can be claimed that the matrix in the right side of Equation (\ref{eqn:coro:2-3}) is also not of FCR. Recall that the matrix $\Psi$ is assumed to be FCR. In addition, from the definition of the matrix $\Psi_{l}^{\perp}$, it is certain that $\Psi_{l}^{\perp} \Psi =0$. It can therefore be declared from Lemma \ref{lemma:2} that the following matrix is not of FCR,
\begin{equation}
\left(\left\{\left[C_{zx} X(\theta) + D_{zu} \Pi\right]^{T}\left[ I-D_{zv} P(\theta)\right]^{-T}\right\}\otimes I\right) \Psi
\label{eqn:coro:2-4}
\end{equation}

Let $\phi$ be a zero column vector having a compatible dimension. With this zero vector $\phi$, Equation (\ref{eqn-app-21}) reduces to the following equation,
\begin{equation}
\hspace*{-1.0cm}\!\!
  \left(\left\{\left[ I-D_{zv} P(\theta)\right]^{-1}\left[C_{zx} X(\theta) + D_{zu} \Pi\right]\right\}^{T}\otimes I\right) \Psi  \delta = 0
\label{eqn-app-23}
\end{equation}

Clearly, when the matrix of Equation (\ref{eqn:coro:2-4}) is not of FCR, there exists at least one nonzero $\delta$ that satisfies this equation, implying that there are more than one values in the set $\mathbf{\Theta}$ that are consistent with the RTIM $\Gamma(\theta)$. Hence the value of the parameter vector $\theta$ can not be determined uniquely. This completes the proof. \hspace{\fill}$\Diamond$

\noindent\textbf{Proof of Theorem \ref{theo:2}:}
From Lemma \ref{lemma:1} and the assumption that $m_{x}\geq m_{\xi}$, it can be claimed that if a RTIM $\Gamma$ is available for System $\mathbf{\Sigma}_{p}$ at the matrix $\Xi$ along the directions of the matrix $\Pi$, then there exist an invertible real matrix $T$, as well as  some other real matrices $G$, $S$, $Z$, $F$, $K$ and $H$, such that
\begin{eqnarray}
& & \hspace*{-1.0cm} \left[\!\!\begin{array}{cc}
				A(\theta) & B(\theta)\\
				C(\theta) & D(\theta)
		\end{array}\!\!\right]
= \left[\!\!\begin{array}{cc}
				T & 0_{m_{x}\times m_{y}}\\
				0_{m_{y}\times m_{x}} & I_{m_{y}}
		\end{array}\!\!\right]
\left[\!\!\begin{array}{cc|c}
\Xi - G\Pi & Z & G \\
-S\Pi & F & S \\
\hline
\Gamma - K \Pi & H & K
\end{array}\!\!\right]\times \nonumber\\
& & \hspace*{3.0cm}
\left[\!\!\begin{array}{cc}
				T^{-1} & 0_{m_{x}\times m_{u}}\\
				0_{m_{u}\times m_{x}} & I_{m_{u}}
		\end{array}\!\!\right]
\label{eqn-app:11}
\end{eqnarray}
Partition the matrix $T$ as $T=[T_{1}\;\; T_{2}]$ in which $T_{1}\in \mathbb{R}^{m_{x}\times m_{\xi}}$ and $T_{2}\in \mathbb{R}^{m_{x}\times (m_{x}-m_{\xi})}$. Then the above equation can be equivalently rewritten as
\begin{eqnarray}
& & \hspace*{-1.25cm} A(\theta)[T_{1}\; T_{2}] \!=\! [T_{1}\; T_{2}] \left[\!\!\!\begin{array}{cc}
\Xi - G\Pi & Z \\
-S\Pi & F
\end{array}\!\!\!\right], \hspace{0.15cm}   B(\theta) \!=\! [T_{1}\; T_{2}]\left[\!\!\!\begin{array}{c}G \\ S \end{array}\!\!\!\right]
\label{eqn-app:12} \\
& & \hspace*{-1.25cm} C(\theta) [T_{1}\;\; T_{2}] =\! [\Gamma - K \Pi \;\; H], \hspace{0.15cm} D(\theta) = K
\label{eqn-app:13}
\end{eqnarray}
Substitute the second equality of Equations (\ref{eqn-app:12}) and (\ref{eqn-app:13}) respectively into its first equality, the following relations are directly obtained,
\begin{equation}
\left[\!\!\begin{array}{cc}
				A(\theta) & B(\theta)\\
				C(\theta) & D(\theta)
		\end{array}\!\!\right] \left[\!\!\begin{array}{c}
				T_{1} \\
				\Pi
		\end{array}\!\!\right]
= \left[\!\!\begin{array}{c}
				T_{1}\Xi \\
				\Gamma
		\end{array}\!\!\right]
\label{eqn-app:14}
\end{equation}

From the construction of the matrix $T_{1}$, it is clear that this matrix is of FCR.

On the contrary, assume that there is a FCR matrix $T_{1}$ that satisfies Equation (\ref{eqn-app:14}). Then there exists at least one matrix $T_{2}\in \mathbb{R}^{m_{x}\times (m_{x}-m_{\xi})}$ such that the matrix $[T_{1}\;\; T_{2}]$ is square and regular. Define matrices $G$, $S$, $Z$, $F$, $K$ and $H$ respectively as
\begin{eqnarray*}
& &\hspace*{-0.1cm} \left[\!\!\begin{array}{c} G \\ S \end{array}\!\!\right]=[T_{1}\;\; T_{2}]^{-1}B(\theta), \hspace{0.5cm}
\left[\!\!\begin{array}{c} Z \\ F \end{array}\!\!\right]=[T_{1}\;\; T_{2}]^{-1}A(\theta)T_{2}
\\
& & \hspace*{-0.1cm} H=C(\theta)T_{2}, \hspace{0.5cm} K=D(\theta)
\end{eqnarray*}
Then direct matrix manipulations show that these matrices with $T=[T_{1}\;\; T_{2}]$ satisfy Equation (\ref{eqn-app:11}), meaning that the associated system matrices $A(\theta)$, $B(\theta)$, $C(\theta)$ and $D(\theta)$ has a parametrization of Lemma \ref{lemma:1}, and is therefore consistent with the RTIM $\Gamma$\footnote[5]{From the definition of a RTIM and Equation (\ref{eqn-app:14}), this conclusion can be directly declared also. The construction of the matrices $T$, $G$, etc., is included here, only for illustrating that the parametrization of \cite{sma2024} is modulo a nonsingular state transformation.}. This completes the proof. \hspace{\fill}$\Diamond$

\noindent\textbf{Proof of Theorem \ref{theo:3}:}
Substitute the system matrix parametrization of Equation (\ref{plant-3}) into Equation (\ref{eqn:pr-1}). The following relation is obtained,
\begin{eqnarray}
\hspace*{-0.5cm} & & \!\!\!\!
        \left[\!\!\begin{array}{c}
				B_{xv}\\
				D_{yv}
        \end{array}\!\! \right]
        \left[ I_{m_{v}} \!\!-\!\! P(\theta)D_{zv}\right]^{-\!1} P(\theta)
        \left[\!\! \begin{array}{cc}
				C_{zx} & D_{zu}
        \end{array}\!\! \right]
\left[\!\!\begin{array}{c}
				T_{1} \\
				\Pi
		\end{array}\!\!\right]  \nonumber\\
\hspace*{-0.5cm} &=&   \!\!\!\!
\left[\!\!\begin{array}{c}
				T_{1}\Xi \\
				\Gamma
		\end{array}\!\!\right] -
\left[\!\!\begin{array}{cc}A_{xx} & B_{xu}\\
				C_{yx} & D_{yu}\\
		\end{array}\!\!\right]
\left[\!\!\begin{array}{c}
				T_{1} \\
				\Pi
		\end{array}\!\!\right]
\label{eqn:pr-2}
\end{eqnarray}
According to Lemma \ref{lemma:3}, there exists a vector $\theta$ such that this equation is satisfied, only if
\begin{equation}
\hspace*{-0.5cm}
        \left[\!\!\begin{array}{c}
				B_{xv}\\
				D_{yv}
        \end{array}\!\! \right]_{l}^{\perp}
\left(
\left[\!\!\begin{array}{c}
				T_{1}\Xi \\
				\Gamma
		\end{array}\!\!\right] -
\left[\!\!\begin{array}{cc}A_{xx} & B_{xu}\\
				C_{yx} & D_{yu}\\
		\end{array}\!\!\right]
\left[\!\!\begin{array}{c}
				T_{1} \\
				\Pi
		\end{array}\!\!\right] \right) =0
\label{eqn:pr-3}
\end{equation}
which can be rewritten equivalently as
\begin{equation}
\hspace*{-0.0cm}
        \left[\!\!\begin{array}{c}
				B_{xv}\\
				D_{yv}
        \end{array}\!\! \right]_{l}^{\perp} \!
\left[\!\!\begin{array}{c}
				T_{1} \\
				0
		\end{array}\!\!\right] \Xi  \!-\!
\left[\!\!\begin{array}{c}
				B_{xv}\\
				D_{yv}
        \end{array}\!\! \right]_{l}^{\perp} \!
\left[\!\!\begin{array}{c} A_{xx} \\
				C_{yx} \\
		\end{array}\!\!\right]T_{1} \!=\!
\left[\!\!\begin{array}{c}
				B_{xv}\\
				D_{yv}
        \end{array}\!\! \right]_{l}^{\perp} \!
\left[\!\!\begin{array}{c }  B_{xu} \Pi  \\
				D_{yu} \Pi - \Gamma
		\end{array}\!\!\right]
\label{eqn:pr-5}
\end{equation}
In addition, when this condition is satisfied, there exists a real matrix $\Omega_{s}$, such that
\begin{eqnarray}
\hspace*{-0.5cm} & & \!\!\!\!
        \left[ I_{m_{v}} \!\!-\!\! P(\theta)D_{zv}\right]^{-\!1} P(\theta)
        \left[\!\! \begin{array}{cc}
				C_{zx} & D_{zu}
        \end{array}\!\! \right]
\left[\!\!\begin{array}{c}
				T_{1} \\
				\Pi
		\end{array}\!\!\right]  \nonumber\\
\hspace*{-0.5cm} &=&   \!\!\!\!
\left[\!\!\begin{array}{c}
				B_{xv}\\
				D_{yv}
        \end{array}\!\! \right]^{\dag}\left( \left[\!\!\begin{array}{c}
				T_{1}\Xi \\
				\Gamma
		\end{array}\!\!\right] -
\left[\!\!\begin{array}{cc}A_{xx} & B_{xu}\\
				C_{yx} & D_{yu}\\
		\end{array}\!\!\right]
\left[\!\!\begin{array}{c}
				T_{1} \\
				\Pi
		\end{array}\!\!\right]\right) +
\left[\!\!\begin{array}{c}
				B_{xv}\\
				D_{yv}
        \end{array}\!\! \right]_{r}^{\perp} \Omega_{s}
\label{eqn:pr-4}
\end{eqnarray}

On the other hand, from the definition of the matrix $\Upsilon_{t}$, as well as that of column vectorization of a matrix, it is obvious that
\begin{equation}
{\rm\bf vec}\left(\left[\!\!\begin{array}{c}
				T_{1} \\
				0
		\end{array}\!\!\right]\right) = \Upsilon_{t} {\rm\bf vec}(T_{1})
\label{eqn:pr-6}
\end{equation}

Vectorize both sides of Equation (\ref{eqn:pr-5}), and substitute the relation of Equation (\ref{eqn:pr-6}) into the vectored equation, the following equation is obtained, directly from the definition of the matrix $\Upsilon_{s}$,
\begin{equation}
\hspace*{-1.0cm}  \Upsilon_{s}  {\rm\bf vec}(T_{1}) \!=\!
{\rm\bf vec}\left(\left[\!\!\begin{array}{c}
				B_{xv}\\
				D_{yv}
        \end{array}\!\! \right]_{l}^{\perp} \!
\left[\!\!\begin{array}{c }  B_{xu} \Pi  \\
				D_{yu} \Pi - \Gamma
		\end{array}\!\!\right]\right)
\label{eqn:pr-7}
\end{equation}
It can therefore be declared from Lemma \ref{lemma:3} that, there exists a vector $\alpha_{t}$, such that
\begin{equation}
\hspace*{-1.0cm}    {\rm\bf vec}(T_{1}) \!=\!
\Upsilon_{s}^{\dag} {\rm\bf vec}\left(\left[\!\!\begin{array}{c}
				B_{xv}\\
				D_{yv}
        \end{array}\!\! \right]_{l}^{\perp} \!
\left[\!\!\begin{array}{c }  B_{xu} \Pi  \\
				D_{yu} \Pi - \Gamma
		\end{array}\!\!\right]\right) +
\Upsilon_{s,r}^{\perp} \alpha_{t}
\label{eqn:pr-8}
\end{equation}

On the basis of the definitions of the matrices $T_{10}$, $T_{1}(\alpha,\Gamma)$ and $\Upsilon_{\alpha,i}|_{i=1}^{m_{\xi}}$, it can be directly claimed from Equation (\ref{eqn:pr-8}) that $T_{1} = T_{1}(\alpha,\Gamma)$, and the matrix $T_{1}$ is of FCR, if and only if
\begin{equation}
\hspace*{0.10cm}  {\rm\bf rank}\left\{ T_{1}(\alpha,\Gamma) \right\} = m_{\xi}
\label{eqn:pr-13}
\end{equation}

On the other hand, the next relation is obtained from Equation (\ref{eqn:pr-4}),
\begin{eqnarray}
& & \hspace*{-0.8cm}\!\!\!\!
        P(\theta)\left\{\! D_{zv}
        \left[\!\!\begin{array}{c}
				B_{xv}\\
				D_{yv}
        \end{array}\!\! \right]^{\dag} \!\left(
  \left[\!\!\begin{array}{c}
				T_{1} \\
				0
		\end{array}\!\!\right] \Xi  \!-\!
  \left[\!\!\begin{array}{c} A_{xx} \\
				C_{yx} \\
		\end{array}\!\!\right]T_{1} \!-\!
  \left[\!\!\begin{array}{c }  B_{xu} \Pi  \\
				D_{yu} \Pi \!-\! \Gamma
		\end{array}\!\!\right] \right) \!+\! \right. \nonumber\\
& & \hspace*{2.6cm} \left. D_{zv}\left[\!\!\begin{array}{c}
				B_{xv}\\
				D_{yv}
        \end{array}\!\! \right]_{r}^{\perp} \!\!\Omega_{s} \!+\! C_{zx}T_{1} \!+\! D_{zu}\Pi \right\}
          \nonumber\\
& & \hspace*{-1.0cm} = \! \left[\!\!\!\!\begin{array}{c}
				B_{xv}\\
				D_{yv}
        \end{array}\!\!\!\! \right]^{\dag} \!\!\! \left(
  \left[\!\!\!\begin{array}{c}
				T_{1} \\
				0
		\end{array}\!\!\!\right] \Xi  \!-\!
  \left[\!\!\!\begin{array}{c} A_{xx} \\
				C_{yx} \\
		\end{array}\!\!\!\right] \!T_{1} \!-\!
  \left[\!\!\!\begin{array}{c }  B_{xu} \Pi  \\
				D_{yu} \Pi \!-\! \Gamma
		\end{array}\!\!\!\right] \!\right) \!+\!
  \left[\!\!\!\begin{array}{c}
				B_{xv}\\
				D_{yv}
        \end{array}\!\!\! \right]_{r}^{\perp}\!\!\! \Omega_{s}
\label{eqn:pr-9}
\end{eqnarray}

Substitute Equation (\ref{plant-4}) into this equation, it can be equivalently rewritten as follows,
\begin{eqnarray}
& & \hspace*{-0.4cm}\!\!\!\!  \sum_{i=1}^{m_{\theta}}\!\theta_{i} \left\{P_{i}\left(D_{zu}\Pi + D_{zv}
\left[\!\!\begin{array}{c}
				B_{xv}\\
				D_{yv}
        \end{array}\!\! \right]^{\dag} \!
  \left[\!\!\begin{array}{c }  -B_{xu} \Pi  \\
				\Gamma \!-\! D_{yu} \Pi
		\end{array}\!\!\right]  \right)\right\} \!+\!
P_{0}C_{zx}T_{1} \!+\nonumber \\
& & \hspace*{-0.4cm}\!\!\!\! (P_{0}D_{zv} \!-\! I) \!
        \left[\!\!\!\begin{array}{c}
				B_{xv}\\
				D_{yv}
        \end{array}\!\!\! \right]^{\dag} \!\!\left(
  \left[\!\!\!\begin{array}{c}
				T_{1} \\
				0
		\end{array}\!\!\!\right] \Xi  \!-\!
  \left[\!\!\!\begin{array}{c} A_{xx} \\
				C_{yx} \\
		\end{array}\!\!\!\right] \! T_{1} \!\right) \!+\! \nonumber \\
& & \hspace*{-0.4cm}\!\!\!\!
\sum_{i=1}^{m_{\theta}}\!\! \left\{\!\! P_{i}C_{zx}(\theta_{i}T_{1}) \!+\! P_{i}D_{zv}\!
        \left[\!\!\!\!\begin{array}{c}
				B_{xv}\\
				D_{yv}
        \end{array}\!\!\!\! \right]^{\dag} \!\!\left(
  \left[\!\!\!\!\begin{array}{c}
				\theta_{i}T_{1} \\
				0
		\end{array}\!\!\!\!\right] \Xi  \!-\!
  \left[\!\!\!\!\begin{array}{c} A_{xx} \\
				C_{yx} \\
		\end{array}\!\!\!\!\right](\theta_{i}T_{1}) \right)\!\right\} \!+\! \nonumber \\
& & \hspace*{-0.4cm}\!\!\!\! (P_{0}D_{zv}-I)
        \left[\!\!\begin{array}{c}
				B_{xv}\\
				D_{yv}
        \end{array}\!\! \right]_{r}^{\perp}\Omega_{s} \!+\!
  \sum_{i=1}^{m_{\theta}}\! \left\{\! P_{i}D_{zv}
        \left[\!\!\begin{array}{c}
				B_{xv}\\
				D_{yv}
        \end{array}\!\! \right]_{r}^{\perp}(\theta_{i}\Omega_{s}) \!\right\}
\nonumber \\
& & \hspace*{-0.7cm}\!\!\!\! =\!  (I-P_{0}D_{zv})\left[\!\!\begin{array}{c}
				B_{xv}\\
				D_{yv}
        \end{array}\!\! \right]^{\dag} \!
  \left[\!\!\begin{array}{c }  -B_{xu} \Pi  \\
				\Gamma \!-\! D_{yu} \Pi
		\end{array}\!\!\right]  \!-\! P_{0}D_{zu}\Pi
\label{eqn:pr-10}
\end{eqnarray}

Denote ${\rm\bf vec}\left(\Omega_{s}\right)$ by $\alpha_{s}$. Let $\alpha_{s,i}$ and $\alpha_{t,i}$ with $i=1, 2, \cdots, m_{\theta}$, be some real vectors that have the same dimension respectively as the vectors $\alpha_{s}$ and $\alpha_{t}$. Then it can be understood without significant difficulties that, $\alpha_{s,i} = {\rm\bf vec}\left(\theta_{i}\Omega_{s}\right)$ and $\alpha_{t,i} = \theta_{i}\alpha_{t}$ are satisfied simultaneously for each $i=1, 2, \cdots, m_{\theta}$, if and only if
\begin{equation}
\hspace*{-1.0cm} {\rm\bf rank}\left(R(\theta,\alpha)\right) = 1
\label{eqn:pr-14}
\end{equation}

Vectorize both sides of Equation (\ref{eqn:pr-10}), and substitute Equation (\ref{eqn:pr-8}) into the vectored equation. Then
from the definitions of the constant vectors $\gamma$ and $w_{i}|_{i=1}^{m_{\theta}}$, as well as those of the constant matrices $W_{t,0}$, $W_{s,0}$, $W_{t,i}|_{i=1}^{m_{\theta}}$ and $W_{s,i}|_{i=1}^{m_{\theta}}$, some straightforward but tedious algebraic manipulations show that, there is a tuple of a vector $\theta$ and a matrix $T_{1}$ with the latter being FCR that satisfies Equation (\ref{eqn:pr-1}), if and only if there is a real vector $\alpha$, together with the same vector $\theta$, such that Equation (\ref{eqn:theo:3-1}) is satisfied under the rank constraints of Equations (\ref{eqn:pr-13}) and (\ref{eqn:pr-14}). This completes the proof. \hspace{\fill}$\Diamond$

\noindent\textbf{Proof of Theorem \ref{theo:5}:}
From the definition of the matrix $R_{in}^{[k]}$, it is obvious that $\sigma_{1}(R^{[k]})$ and $\sigma_{i}(R^{[k]}) - \lambda_{2}$ with $1\leq i\leq \tau$ are singular values of the matrix $R_{in}^{[k]}$, while $u_{i}^{[k]}$ and $v_{i}^{[k]}$ are respectively left and right singular vectors associated with its $i$th singular value. That is, Equation (\ref{eqn:theo:5-1}) is actually a SVD for the matrix $R_{in}^{[k]}$.

Denote an SVD of the matrix $R_{in}$ by $U_{in}\Sigma_{in}V_{in}^{T}$. Then it can be proved that the subdifferential of the cost function $\overline{J}^{[k]}(R_{in})$ at $R_{in}$ has the following expression\cite{rfp2010,tbls2024},
\begin{equation}
\partial \overline{J}^{[k]}(R_{in}) \!=\! \left\{\!\!\! \begin{array}{c}  R_{in} \!-\! R^{[k]}  \!+\! \lambda_{2} \left[U_{in}V_{in}^{T} \!+\! W \!-\! u_{1}^{[k]}v_{1}^{[k],T}\right]\\
{\rm with}\;\; U_{in}^{T}W=0,\;\; WV_{in}=0,\;\; ||W||_{2}\leq 1 \end{array} \!\!\!\right\}
\label{eqn:ns-6}
\end{equation}

Define matrices $U_{in}^{[k]}$, $V_{in}^{[k]}$ and $W^{[k]}$ respectively as
\begin{eqnarray*}
& &
U_{in}^{[k]} \!=\! {\rm\bf row}\{u_{i}^{[k]}|^{\tau}_{i=1}\}, \hspace{0.25cm}
V_{in}^{[k]} \!=\! {\rm\bf row}\{v_{i}^{[k]}|^{\tau}_{i=1}\} \\
& &
W^{[k]} \!=\! \sum_{i=\tau+1}^{m_{\theta}+1}\!\frac{\sigma_{i}(R^{[k]})}{\lambda_{2}}u_{i}^{[k]}v_{i}^{[k],T}
\end{eqnarray*}
Then it is obvious that $||W^{[k]}||_{2}<1$, and
\begin{eqnarray}
& & \hspace*{-0.5cm}
U_{in}^{[k],T}W^{[k]}=0,\;\; W^{[k]}V_{in}^{[k]}=0
\label{eqn:ns-7}   \\
& & \hspace*{-0.5cm}
R_{in}^{[k]} \!-\! R^{[k]}  \!+\! \lambda_{2} \left[U_{in}^{[k]}V_{in}^{[k],T} \!+\! W^{[k]} \!-\! u_{1}^{[k]}v_{1}^{[k],T}\right] \!=\!0
\label{eqn:ns-8}
\end{eqnarray}
It can therefore be declared from Equation (\ref{eqn:ns-6}) that $0 \in \partial \overline{J}^{[k]}(R_{in}^{[k]})$. Recall that the cost function $\overline{J}^{[k]}(R_{in})$ is strictly convex. From convex optimizations \cite{hl2001}, it can be further claimed that $R_{in}^{[k]}$ is the unique global optimizer of this cost function.
\hspace{\fill}$\Diamond$

\noindent\textbf{Proof of Theorem \ref{theo:4}:}
Denote respectively by vectors $\widehat{\theta}$ and $\widehat{\alpha}$ the optimal value of the vectors $\overline{\theta}$ and $\overline{\alpha}$, that minimize the cost function of Equation (\ref{eqn:pr-28}) under the required rank constraints. That is,
\begin{eqnarray}
& & \left(\widehat{\theta}, \widehat{\alpha}\right) = arg\;\min_{\overline{\theta},\; \overline{\alpha}}\left|\left|e(\overline{\theta},  \overline{\alpha}, \widehat{\Gamma})\right|\right| \nonumber \\
& & \hspace*{1.0cm} subject\;\; to\;\; \overline{\theta} \in \mathbf{\Theta},\;\;
     {\rm\bf rank}\left(R(\overline{\theta}, \overline{\alpha})\right)=1    \label{eqn-app-26} \\
& & \hspace*{2.55cm}
     {\rm\bf rank}\left(T_{1}(\overline{\theta}, \overline{\alpha}, \widehat{\Gamma})\right)= m_{\xi} \nonumber
\end{eqnarray}
Moreover, denote matrices $T_{1}(\widehat{\theta}, \widehat{\alpha}, \widehat{\Gamma})$ and
$C(\widehat{\theta})T_{1}(\widehat{\theta}, \widehat{\alpha}, \widehat{\Gamma}) + D(\widehat{\theta}) \Pi$ respectively
by matrices $\widehat{T}_{1}$ and $\widetilde{\Gamma}$. Then the following equality can be established from the definition of the minimization problem,
\begin{equation}
\left[\!\!\begin{array}{cc}
				A(\widehat{\theta}) & B(\widehat{\theta})\\
				C(\widehat{\theta}) & D(\widehat{\theta})
		\end{array}\!\!\right] \left[\!\!\begin{array}{c}
				\widehat{T}_{1} \\
				\Pi
		\end{array}\!\!\right]
= \left[\!\!\begin{array}{c}
				\widehat{T}_{1}\Xi \\
				\widetilde{\Gamma}
		\end{array}\!\!\right]
\label{eqn:pr-15}
\end{equation}

On the other hand, from the definitions of the vector valued function $e(\overline{\theta}, \overline{\alpha}, \widehat{\Gamma})$, it is obvious that
\begin{equation}
e({\theta},  {\alpha}, \widehat{\Gamma}) \!-\! e({\theta},  {\alpha}, {\Gamma})
\!=\! \!
\sum_{i=1}^{m_{\theta}} \! {\theta}_{i} \left[w_{i}(\widehat{\Gamma}) \!-\!w_{i}({\Gamma})\right]
\!-\! \gamma(\widehat{\Gamma})
\!+\! \gamma({\Gamma})
\label{eqn-app-28}
\end{equation}
In addition, from the definitions of the vectors $\gamma(\Gamma)$ and $w_{i}(\Gamma)$, in which $i=1, 2, \cdots, m_{\theta}$, that are given immediately before Theorem \ref{theo:3}, the following equalities can be straightforwardly established,
\begin{equation}
\hspace*{-0.2cm}  \gamma(\widehat{\Gamma}) \!-\! \gamma({\Gamma})  \!=\! F_{\gamma} {\rm\bf vec}\!\!\left(\widehat{\Gamma} \!-\! \Gamma \right), \hspace{0.15cm}
w_{i}(\widehat{\Gamma}) \!-\! w_{i}(\Gamma) \!=\! F_{w,i} {\rm\bf vec}\!\!\left(\widehat{\Gamma} \!-\! \Gamma \right)
\label{eqn-app-27}
\end{equation}
in which
{\footnotesize \begin{eqnarray*}
& & \hspace*{-0.7cm}  F_{\gamma}  \!=\!  I \!\otimes\! \left(\!\!\!(I \!-\! P_{0}D_{zv})\! \left[\!\!\!\begin{array}{c}
				B_{xv}\\
				D_{yv}
        \end{array}\!\!\! \right]^{\dag} \!\!
  \left[\!\!\!\begin{array}{c }  0  \\
				I
		\end{array}\!\!\!\right]\!\!\right) \!-\!
  \left\{\! I \!\otimes\! (P_{0}C_{zx}) \!+\! \left(\! I \!\otimes\! \left[\!(P_{0}D_{zv} \!-\! I)\!\times
        \begin{array}{c}
				\;\; \\
				\;\;
        \end{array} \right.\right. \right.  \\
& & \hspace*{-0.2cm}
     \left. \left.\left.\left[\!\!\!\begin{array}{c}
				B_{xv}\\
				D_{yv}
        \end{array}\!\!\! \right]^{\dag}\right]\right) \!\left(\! \left(\Xi^{T} \!\!\otimes\! I \!\right) \! \Upsilon_{t} \!-\!  I \!\otimes\!
  \left[\!\!\begin{array}{c} A_{xx} \\
				C_{yx} \\
		\end{array}\!\!\right] \right) \right\} \! \Upsilon_{s}^{\dag}\!
         \left[I \!\otimes\! \left(\left[\!\!\!\!\begin{array}{c}
				B_{xv}\\
				D_{yv}
        \end{array}\!\!\!\! \right]_{l}^{\perp} \!
  \left[\!\!\!\!\begin{array}{c }  0  \\
				-I
		\end{array}\!\!\!\!\right]\!\!\right) \!\!\right\}  \\
& & \hspace*{-0.7cm} F_{w,i} \!=\! I \!\otimes\!\! \left( \!\!P_{i}D_{zv} \!\!\left[\!\!\!\!\begin{array}{c}
				B_{xv}\\
				D_{yv}
        \end{array}\!\!\!\! \right]^{\dag} \!
        \left[\!\!\!\begin{array}{c }  0 \\
				I
		\end{array}\!\!\!\right] \!\!\right) \!+\!
  \left\{\!\! I \!\otimes\! (P_{i}C_{zx}) \!+\!
   \left[\! I \!\otimes\! \left( \!\! P_{i}D_{zv}\!
        \left[\!\!\!\!\begin{array}{c}
				B_{xv}\\
				D_{yv}
        \end{array}\!\!\!\! \right]_{r}^{\perp} \!\!\right)\!\!\right] \!\times \right.  \\
& & \hspace*{0.6cm}
        \left. \left(\! \left(\! \Xi^{T} \!\otimes\! I\right) \!\Upsilon_{t} \!-\! I \!\otimes\! \left[\!\!\!\!\begin{array}{c} A_{xx} \\
				C_{yx} \\
		\end{array}\!\!\!\!\right] \!\right) \!\!\right\} \!\Upsilon_{s}^{\dag} \!
  \left[I \!\otimes\! \left( \!\left[\!\!\!\!\begin{array}{c}
				B_{xv}\\
				D_{yv}
        \end{array}\!\!\!\! \right]_{l}^{\perp} \!\!
  \left[\!\!\!\!\begin{array}{c }  0  \\
				-I
		\end{array}\!\!\!\!\right] \!\right) \!\right]
\end{eqnarray*}}
Here, the identity matrices in general have different dimensions, which are omitted for space considerations. Obviously, all the matrices $F_{\gamma}$ and $F_{w,i}|_{i=1}^{m_{\theta}}$ do not depend on either the value of the parameter vector $\theta$, or the value of the RTIM $\Gamma$, and each of their elements takes a value with a finite magnitude.

Recall that the matrix $\Gamma$ is the actual value of the RTIM of System $\mathbf{\Sigma}_{p}(\theta)$ at the matrix $\Xi$ along the directions of the matrix $\Pi$. It is obvious that
\begin{equation}
e({\theta},  {\alpha}, {\Gamma}) =0, \hspace{0.50cm}
\left|\left|e({\theta},  {\alpha}, \widehat{\Gamma})\right|\right| \geq
\left|\left|e(\widehat{\theta},  \widehat{\alpha}, \widehat{\Gamma})\right|\right|
\label{eqn-app-25}
\end{equation}
Moreover, when the parameter vector $\theta$ is recoverable, $e(\overline{\theta},  \overline{\alpha}, {\Gamma}) =0$ only when $\overline{\theta} = \theta$.

Substitute Equations (\ref{eqn-app-28}) and (\ref{eqn-app-27}) into Equation (\ref{eqn-app-25}). The next inequality is obtained directly from the definition of the induced norm of a matrix \cite{gv1989,hj1991},
\begin{eqnarray}
\left|\left|e(\widehat{\theta},  \widehat{\alpha}, \widehat{\Gamma})\right|\right| &\leq&
\left|\left|\left(\sum_{i=1}^{m_{\theta}} \! {\theta}_{i}F_{w,i} \!-\! F_{\gamma}\right)
{\rm\bf vec}\left(\widehat{\Gamma} \!-\! \Gamma \right)\right|\right| \nonumber \\
&\leq& \left|\left|\sum_{i=1}^{m_{\theta}} \! {\theta}_{i}F_{w,i} \!-\! F_{\gamma}\right|\right| \!\times\!
\left|\left|{\rm\bf vec}\left(\widehat{\Gamma} \!-\! \Gamma \right)\right|\right|
\label{eqn-app-29}
\end{eqnarray}

Partition the vector $e(\widehat{\theta},  \widehat{\alpha}, \widehat{\Gamma})$ into $m_{v}$ row blocks with each block has $m_{\xi}$ rows. Denote the $i$th block counting from the ceiling as $e_{i}(\widehat{\theta},  \widehat{\alpha}, \widehat{\Gamma})$ with $i=1,2,\cdots,m_{v}$. Define a matrix $E(\widehat{\theta},  \widehat{\alpha}, \widehat{\Gamma})$ as
\begin{displaymath}
E(\widehat{\theta},  \widehat{\alpha}, \widehat{\Gamma}) \!=\! \left[e_{1}(\widehat{\theta},  \widehat{\alpha}, \widehat{\Gamma})\;\;\;\; e_{2}(\widehat{\theta},  \widehat{\alpha}, \widehat{\Gamma})\;\;\;\; \cdots\;\;\;\; e_{m_{v}}(\widehat{\theta},  \widehat{\alpha}, \widehat{\Gamma})\right]
\end{displaymath}
Then from the definition of the vector valued function $e({\theta},  {\alpha}, {\Gamma})$, direct but tedious algebraic manipulations show that
\begin{equation}
  \left[\!\!\!\begin{array}{cc}
				A(\widehat{\theta}) & B(\widehat{\theta})\\
				C(\widehat{\theta}) & D(\widehat{\theta})
		\end{array}\!\!\!\right] \! \left[\!\!\!\begin{array}{c}
				\widehat{T}_{1} \\
				\Pi
		\end{array}\!\!\!\right]
\!=\! \left[\!\!\!\begin{array}{c}
				\widehat{T}_{1}\Xi \\
				\widehat{\Gamma}
		\end{array}\!\!\!\right] \!+\!
  \left[\!\!\!\begin{array}{c}
				B_{xv}\\
				D_{yv}
        \end{array} \!\!\!\right]
        \left[ I \!-\! P(\widehat{\theta})D_{zv} \!\right]^{-1} \!\! E(\widehat{\theta},  \widehat{\alpha}, \widehat{\Gamma})
\label{eqn:pr-15-x}
\end{equation}

Compare Equation (\ref{eqn:pr-15-x}) with Equation (\ref{eqn:pr-15}), it is clear that
\begin{equation}
  \left[\!\!\begin{array}{c}
				B_{xv}\\
				D_{yv}
        \end{array}\!\!\right]
        \left[ I \!-\! P(\widehat{\theta})D_{zv} \right]^{-1} E(\widehat{\theta},  \widehat{\alpha}, \widehat{\Gamma})
=
   \left[\!\!\begin{array}{c}
				0 \\
				\widetilde{\Gamma} - \widehat{\Gamma}
		\end{array}\!\!\right]
\label{eqn-app-30}
\end{equation}
Hence, from the construction of the matrix $E(\widehat{\theta},  \widehat{\alpha}, \widehat{\Gamma})$, it is immediate that the following equality is valid,
\begin{equation}
  {\rm\bf vec}\left(\widetilde{\Gamma} - \widehat{\Gamma}\right)
= \left\{I \otimes \left(D_{yv} \left[ I \!-\! P(\widehat{\theta})D_{zv} \right]^{-1}\right)\right\} e(\widehat{\theta},  \widehat{\alpha}, \widehat{\Gamma})
\label{eqn-app-31}
\end{equation}

Denote $\widehat{\theta} - \theta$, $P(\widehat{\theta}) - P(\theta)$, $\widetilde{\Gamma} - \Gamma$ and $\widehat{T}_{1}-T_{1}$ respectively by $\delta_{\theta}$, $\Delta_{P}$, $\Delta_{\Gamma}$ and $\Delta_{T_{1}}$. Moreover, denote ${\rm\bf vec}\left(\Delta_{\Gamma}\right)$, ${\rm\bf vec}\left(\Delta_{P}\right)$ and ${\rm\bf vec}\left(\Delta_{T_{1}}\right)$ respectively by $\delta_{\Gamma}$, $\delta_{P}$ and $\delta_{T_{1}}$. From the definitions of the matrices $P(\theta)$ and $\Psi$, that are given respectively by Equations (\ref{plant-4}) and (\ref{eqn:par-vec-1}), it is immediate that
\begin{equation}
\delta_{P} = \Psi \delta_{\theta}
\label{eqn:pr-20}
\end{equation}
Moreover, from the definitions of the matrices $\widetilde{\Gamma}$, $\Gamma$ and $\widehat{\Gamma}$, it can be claimed that the equality
$\widetilde{\Gamma} - \Gamma = \left(\widetilde{\Gamma} - \widehat{\Gamma}\right)+ \left(\widehat{\Gamma} - \Gamma\right)$ is well defined. In addition, from Equations (\ref{eqn-app-29}) and (\ref{eqn-app-31}), as well as the well-known triangular inequality for a vector norm \cite{gv1989,hj1991,zyl2018}, the following inequalities can be established,
{\footnotesize \begin{eqnarray}
& & \hspace*{-1.2cm}   \left|\left|\delta_{\Gamma}\right|\right|  \leq
  \left|\left|{\rm\bf vec}\left(\widetilde{\Gamma} - \widehat{\Gamma} \right)\right|\right| +
  \left|\left|{\rm\bf vec}\left(\widehat{\Gamma} \!-\! \Gamma \right)\right|\right|   \nonumber \\
& & \hspace*{-1.0cm}\leq
  \left|\left|I \!\otimes\! \left(D_{yv} \left[ I \!-\! P(\widehat{\theta})D_{zv} \right]^{-1}\!\right)\right|\right|\!\times\!
  \left|\left|e(\widehat{\theta},  \widehat{\alpha}, \widehat{\Gamma})\right|\right| \!+\!
  \left|\left|{\rm\bf vec}\left(\widehat{\Gamma} \!-\! \Gamma \right)\right|\right| \nonumber \\
& & \hspace*{-1.0cm} \leq \!\!
  \left\{\!\! 1 \!+\!\! \left|\!\left|\! I \!\otimes\!\! \left( \! D_{yv} \!\!\left[ I \!-\! P(\widehat{\theta})D_{zv} \! \right]^{\!-\!1} \!\right)\! \!\right|\!\right| \!\times\!
  \left|\!\left|\! \sum_{i=1}^{m_{\theta}} \!\! {\theta}_{i}F_{w,i} \!\!-\! F_{\gamma} \!\right|\!\right| \!\right\} \!\!
\left|\!\left|\!{\rm\bf vec} \!\left(\!\widehat{\Gamma} \!-\! \Gamma \! \right)\!\right|\!\right|
\label{eqn-app-32}
\end{eqnarray}}

Note that invertibility of the matrix $I \!-\! P(\widehat{\theta})D_{zv} $ is guaranteed by Assumption \ref{assum:1}. The last inequality in the above equation imply that if $\widehat{\Gamma}$ is an sufficiently accurate estimate for the RTIM $\Gamma$, then through optimizing the cost function of Equation (\ref{eqn:pr-28}), the suggested recovery procedure can provide an estimate for the parameter vector $\theta$ that makes $\Delta_{\Gamma}$, that is, $\widetilde{\Gamma} \!-\! \Gamma$, close to a zero matrix, provided that this parameter vector is recoverable.

Note also that
\begin{eqnarray*}
& & \hspace*{-0.9cm} [I-P(\widehat{\theta})D_{zv}]^{-1}P(\widehat{\theta}) = [I-P(\theta)D_{zv}]^{-1}P(\theta) + \\
& & \hspace*{2.1cm} [I-P(\theta)D_{zv}]^{-1} \Delta_{P} [I-D_{zv}P(\widehat{\theta})]^{-1}
\end{eqnarray*}
Substitute this relation and Equation (\ref{plant-3}) into Equations (\ref{eqn:pr-1}) and (\ref{eqn:pr-15}). Straightforward algebraic manipulations show that
\begin{eqnarray}
& & \hspace*{-0.85cm}
        \left[\!\!\begin{array}{c}
				B_{xv}\\
				D_{yv}
        \end{array}\!\!\right]
        \left[ I \!-\! P(\theta)D_{zv} \right]^{-1} \!\Delta_{P}
        \left[ I \!-\! D_{zv} P(\widehat{\theta}) \right]^{-1} \!\!
        \left[ C_{zx} T_{1} \!+\!  D_{zu} \Pi \right] \!+  \nonumber\\
& & \hspace*{-0.85cm}
        \left[\!\!\begin{array}{c}
				B_{xv}\\
				D_{yv}
        \end{array}\!\!\right]
        \left[ I - P(\theta)D_{zv} \right]^{-1} P(\theta) C_{zx} \Delta_{T_{1}} +  \nonumber\\
& & \hspace*{-0.85cm}
        \left[\!\!\begin{array}{c}
				B_{xv}\\
				D_{yv}
        \end{array}\!\!\right]
        \left[ I - P(\theta)D_{zv} \right]^{-1} \Delta_{P}
        \left[ I - D_{zv} P(\widehat{\theta}) \right]^{-1} C_{zx} \Delta_{T_{1}}
         \nonumber\\
& & \hspace*{-1.0cm} =
        \left[ \!\!\begin{array}{c}
				\Delta_{T_{1}} \Xi\\
				\Delta_{\Gamma}
        \end{array}\!\! \right] -
        \left[\!\!\begin{array}{c}
				A_{xx}\\
				C_{yx}
        \end{array}\!\! \right] \Delta_{T_{1}}
\label{eqn:pr-16}
\end{eqnarray}

On the other hand, direct matrix manipulations show that
\begin{eqnarray*}
& & \hspace*{-0.6cm} [I \!-\! D_{zv}P(\widehat{\theta})]^{-1} \!=\! [I \!-\! D_{zv}P(\theta)]^{-1} \!+\! [I \!-\! D_{zv}P(\theta)]^{-1}D_{zv}\!\times \\
& & \hspace*{4.5cm} \Delta_{P} [I-D_{zv}P(\widehat{\theta})]^{-1}
\end{eqnarray*}
On the basis of this relation, the following equality is obtained from Equation (\ref{eqn:pr-16}),
\begin{eqnarray}
& & \hspace*{-1.00cm}
        \left[\!\!\begin{array}{c}
				B_{xv}\\
				D_{yv}
        \end{array}\!\!\right]
        \left[ I \!-\! P(\theta)D_{zv} \right]^{-1} \!\Delta_{P}
        \left[ I \!-\! D_{zv} P(\theta) \right]^{-1} \!\!
        \left[ C_{zx} T_{1} \!+\!  D_{zu} \Pi \right] \!+  \nonumber\\
& & \hspace*{-1.00cm}
        \left[\!\!\begin{array}{c}
				B_{xv}\\
				D_{yv}
        \end{array}\!\!\right]
        \left[ I \!-\! P(\theta)D_{zv} \right]^{-1} \!\!P(\theta) C_{zx} \Delta_{T_{1}} \!+\!
        \left[\!\!\begin{array}{c}
				A_{xx}\\
				C_{yx}
        \end{array}\!\! \right] \Delta_{T_{1}} \!-\!
        \left[ \!\!\begin{array}{c}
				\Delta_{T_{1}} \Xi\\
				 0
        \end{array}\!\! \right]
         \nonumber\\
& & \hspace*{-1.3cm} =
        \left[ \!\!\begin{array}{c}
				0 \\
				\Delta_{\Gamma}
        \end{array}\!\! \right] \!-\!
        \left[\!\!\begin{array}{c}
				B_{xv}\\
				D_{yv}
        \end{array}\!\!\right]
        \left[ I - P(\theta)D_{zv} \right]^{-1} \Delta_{P}
        \left[ I - D_{zv} P(\widehat{\theta}) \right]^{-1} C_{zx} \Delta_{T_{1}} \!-\!  \nonumber \\
& & \hspace*{-1.00cm}
        \left[\!\!\begin{array}{c}
				B_{xv}\\
				D_{yv}
        \end{array}\!\!\right]
        \left[ I \!-\! P(\theta)D_{zv} \right]^{-1} \!\Delta_{P}
        \left[ I \!-\! D_{zv} P(\theta) \right]^{-1} \!\! D_{zv} \times \nonumber \\
& & \hspace*{2.00cm}
        \Delta_{P}  \left[ I \!-\! D_{zv} P(\widehat{\theta}) \right]^{-1}
        \left[ C_{zx} T_{1} \!+\!  D_{zu} \Pi \right]
\label{eqn:pr-17}
\end{eqnarray}
That is,
\begin{eqnarray}
& & \hspace*{-1.00cm}
        B_{xv} \left[ I \!-\! P(\theta)D_{zv} \right]^{-1} \!\Delta_{P}
        \left[ I \!-\! D_{zv} P(\theta) \right]^{-1} \!\!
        \left[ C_{zx} T_{1} \!+\!  D_{zu} \Pi \right] \!+  \nonumber\\
& & \hspace*{-1.00cm}
        \left(B_{xv}  \left[ I \!-\! P(\theta)D_{zv} \right]^{-1} \!\!P(\theta) C_{zx}  \!+\!
        A_{xx}\right) \Delta_{T_{1}} \!-\! \Delta_{T_{1}} \Xi
         \nonumber\\
& & \hspace*{-1.3cm} = o\!\left( ||\Delta_{P}||, ||\Delta_{T_{1}}||  \right)
\label{eqn:pr-18}
\end{eqnarray}
while
\begin{eqnarray}
& & \hspace*{-1.00cm}
        D_{yv} \left[ I \!-\! P(\theta)D_{zv} \right]^{-1} \!\Delta_{P}
        \left[ I \!-\! D_{zv} P(\theta) \right]^{-1} \!\!
        \left[ C_{zx} T_{1} \!+\!  D_{zu} \Pi \right] \!+  \nonumber\\
& & \hspace*{-1.00cm}
		\left(D_{yv} \left[ I \!-\! P(\theta)D_{zv} \right]^{-1} \!\!P(\theta) C_{zx}  \!+\!
    	C_{yx} \right) \Delta_{T_{1}}         \nonumber\\
& & \hspace*{-1.3cm} =  \Delta_{\Gamma} + o\!\left( ||\Delta_{P}||, ||\Delta_{T_{1}}||  \right)
\label{eqn:pr-19}
\end{eqnarray}

Vectorize both sides of Equations (\ref{eqn:pr-18}) and (\ref{eqn:pr-19}). Note that different norms of a finite dimensional matrix are equivalent to each other \cite{gv1989,hj1991}. Using the defined symbols $\delta_{\Gamma}$, $\delta_{T_{1}}$ and $\delta_{P}$, the following relations are derived from Equations (\ref{eqn:pr-18}) and (\ref{eqn:pr-19}).
\begin{eqnarray}
& & \hspace*{-0.90cm}
                 R_{xv}(\theta)\delta_{\theta} \!+\!  R_{xx}(\theta) \delta_{T_{1}}
          \!=\! o\!\left( ||\delta_{\theta}||, ||\delta_{T_{1}}||  \right)
\label{eqn:pr-21}  \\
& & \hspace*{-0.90cm}
                R_{yv}(\theta)\delta_{\theta} \!+\!  R_{yx}(\theta) \delta_{T_{1}}
          \!=\! \delta_{\Gamma} + o\!\left( ||\delta_{\theta}||, ||\delta_{T_{1}}|| \right)
\label{eqn:pr-22}
\end{eqnarray}

From Lemma \ref{lemma:3} and Equation (\ref{eqn:pr-21}), it can be claimed that there exists a vector $\delta_{\alpha}$, such that
\begin{equation}
\hspace*{-0.00cm} \left[\!\!\!\begin{array}{c} \delta_{\theta} \\ \delta_{T_{1}} \end{array}\!\!\!\right] \!=\!
                 \left[ R_{xv}(\theta)\;\;   R_{xx}(\theta)\right]_{r}^{\perp} \delta_{\alpha} +
             o\!\left( ||\delta_{\theta}||, ||\delta_{T_{1}}||  \right)
\label{eqn:pr-23}  \\
\end{equation}
Substitute it into Equation (\ref{eqn:pr-22}). The following equation is obtained.
\begin{equation}
\hspace*{-0.00cm}
                \left[ R_{yv}(\theta)\;\;  R_{yx}(\theta) \right] \left[ R_{xv}(\theta)\;\;   R_{xx}(\theta)\right]_{r}^{\perp} \delta_{\alpha}  \!= \! \delta_{\Gamma} + o\!\left( ||\delta_{\theta}||, ||\delta_{T_{1}}||  \right)
\label{eqn:pr-24}
\end{equation}
Then when the matrix $\left[ R_{yv}(\theta)\;\;  R_{yx}(\theta) \right] \left[ R_{xv}(\theta)\;\;   R_{xx}(\theta)\right]_{r}^{\perp}$ is of FCR, the next expression can be obtained for $\delta_{\alpha}$ from Lemma \ref{lemma:3},
\begin{equation}
\hspace*{-0.00cm}
                 \delta_{\alpha}  \!=\! \left(\left[ R_{yv}(\theta)\;\;  R_{yx}(\theta) \right] \left[ R_{xv}(\theta)\;\;   R_{xx}(\theta)\right]_{r}^{\perp}\right)^{\dag} \!\!\delta_{\Gamma} \!+\! o\!\left( ||\delta_{\theta}||, ||\delta_{T_{1}}||\right)
\label{eqn:pr-25}
\end{equation}

Substitute Equation (\ref{eqn:pr-25}) back into Equation (\ref{eqn:pr-23}), a relation is obtained between the error vectors $\delta_{\theta}$ and $\delta_{\Gamma}$,
\begin{eqnarray}
& & \hspace*{-1.00cm}
                 \delta_{\theta}  \!=\! \left[ I_{m_{\theta}}\;\; 0\right] \!\left[ R_{xv}(\theta)\;\;   R_{xx}(\theta)\right]_{r}^{\perp} \! \left(\left[ R_{yv}(\theta)\;\;  R_{yx}(\theta) \right] \left[ R_{xv}(\theta)\;\;   R_{xx}(\theta)\right]_{r}^{\perp}\right)^{\dag} \!\!\delta_{\Gamma} \!+\!  \nonumber\\
& & \hspace*{3.00cm}
                 o\!\left( ||\delta_{\theta}||, ||\delta_{T_{1}}||\right)
\label{eqn:pr-26}
\end{eqnarray}
meaning that a small $\delta_{\Gamma}$ does not lead to a large $\delta_{\theta}$. Therefore, the suggested recovery procedure is robust against errors in the RTIM $\Gamma$.

On the contrary, assume that the matrix $\left[ R_{yv}(\theta)\;\;  R_{yx}(\theta) \right]\times$ $\left[ R_{xv}(\theta)\;\;   R_{xx}(\theta)\right]_{r}^{\perp}$ is not of FCR. Then its right null space is not trivial. Moreover, according to Equation (\ref{eqn:pr-24}) and Lemma \ref{lemma:3}, there is a vector $\beta$ whose magnitude can be arbitrarily large, such that
\begin{eqnarray}
& & \hspace*{-1.00cm}
                 \delta_{\alpha}  \!=\! \left(\left[ R_{yv}(\theta)\;\;  R_{yx}(\theta) \right] \left[ R_{xv}(\theta)\;\;   R_{xx}(\theta)\right]_{r}^{\perp}\right)^{\dag} \!\!\delta_{\Gamma} \!+\! \nonumber\\
& & \hspace*{-0.40cm} \left(\!\left[ R_{yv}(\theta)\;\:  R_{yx}(\theta) \right] \left[ R_{xv}(\theta)\;\:   R_{xx}(\theta)\right]_{r}^{\perp}\!\!\right)_{r}^{\perp} \!\!\!\beta \!+\!
o\!\left(\! |\!|\delta_{\theta}|\!|, |\!|\delta_{T_{1}}|\!|\!\right)
\label{eqn:pr-27}
\end{eqnarray}

Note that both the matrices $\left[ R_{xv}(\theta)\;\;   R_{xx}(\theta)\right]_{r}^{\perp}$ and $\left(\left[ R_{yv}(\theta)\;\;  R_{yx}(\theta) \right]\times\right.$ $\left. \left[ R_{xv}(\theta)\;\;   R_{xx}(\theta)\right]_{r}^{\perp}\right)_{r}^{\perp}$ are of FCR. Their product is therefore also of FCR. Hence it is no longer valid to get a relation between $\delta_{\theta}$ and $\delta_{\Gamma}$ through substituting Equation (\ref{eqn:pr-27}) back into Equation (\ref{eqn:pr-23}), recalling that no constraints are put on the magnitude of an element of the vector $\beta$, that makes $||\delta_{\alpha}||$ in an order different from $||\delta_{\Gamma}||$. This means that there may exists an error vector $\delta_{\Gamma}$ with each of its elements having a very small magnitude, that leads to a $\delta_{\theta}$ with some of its elements having a very large magnitude. That is, the suggested estimation procedure is very sensitive to some estimation errors of the RTIM $\Gamma$, and hence not robust against them.

This completes the proof. \hspace{\fill}$\Diamond$

\vspace{-0.2cm}

\section*{\small \hspace*{-0.0cm} Appendix II: System Matrices for the Simulation Example}
\vspace{-0.1cm}

\begin{eqnarray*}
& & \hspace*{-0.70cm} A_{xx} \!=\!\left[\!\! \begin{array}{cccc}
  -r_{p,1} & r_{p,1} \!-\! r_{z} & 0                & 0 \\
  0      & -r_{p,2}        & 0                & 0 \\
  0      & 0             & 0                & 1 \\
  -1     & 1             & 0  & 0  \end{array}  \!\!\right]\!,\hspace{0.2cm}
B_{xu} \!=\! \left[\!\! \begin{array}{c}
  0 \\
  k \\
  0 \\
  0 \end{array} \!\!  \right]\!,\hspace{0.20cm}
  B_{xv}=\left[\!\! \begin{array}{cc}
  0 & 0 \\
  0 & 0 \\
  0 & 0 \\
  1 & 0 \end{array} \!\!  \right]  \\
& & \hspace*{-0.70cm}
  C_{zx}=\left[\!\! \begin{array}{cccc}
  0 & 0 & 0 & -2 \\
  0 & 0 & 0 & 0  \\
  0 & 0 & -1 & 0  \end{array} \!\!  \right],\hspace{0.25cm}
  D_{zv}=\left[\!\! \begin{array}{cc}
  0 & 0 \\
  0 & 1 \\
  0 & 0 \end{array} \!\!  \right],\hspace{0.25cm}	
  D_{zu}=\left[\!\! \begin{array}{c}
  0 \\
  0 \\
  0 \end{array} \!\!  \right]  \\
& & \hspace*{-0.70cm} C_{yx}=\left[\!\! \begin{array}{cccc}
  -1 & 1 & \omega_{z}^{2} &
  2\zeta_{z}\omega_{z} \end{array} \!\!  \right],\hspace{0.25cm} D_{yu}=0, \hspace{0.25cm}
  D_{yv}=[1\;\; 0]  \\
& & \hspace*{-0.70cm}
P_{1}=\left[\!\! \begin{array}{ccc}
  0 & 0 & 0       \\
  1 & 0 & 0 \end{array} \!\!  \right] ,\hspace{0.25cm} P_{2}= \left[\!\! \begin{array}{ccc}
  0 & 1 & 0       \\
  0 & 0 & 1 \end{array} \!\!  \right]
\end{eqnarray*}


\end{document}